\documentclass[twocolumn,dvipsnames,twocolappendix]{aastex63}

\usepackage{graphicx}
\usepackage{amsmath,mathtools} \usepackage{amsfonts} \usepackage{amssymb}
\usepackage{placeins}
\usepackage{hyperref}
\usepackage{multirow}
\usepackage[T1]{fontenc}
\usepackage[utf8]{inputenc}
\usepackage{CJK}

\renewcommand{\d}{\mathinner{{\rm d}\!}}
\renewcommand{\vec}[1]{\mathinner{{\mathbf{#1}}}}

\newcommand{\Msun}{\mathinner{{\rm M}_{\odot}}}

\newcommand{\kpc}{\mathinner{{\rm kpc}}}
\newcommand{\pc}{\mathinner{{\rm pc}}}
\newcommand{\Gyr}{\mathinner{{\rm Gyr}}}
\newcommand{\Myr}{\mathinner{{\rm Myr}}}

\defcitealias{Gobat2016}{GH16}
\defcitealias{Kroupa1993}{KTG93}

\begin{document}

\title{Panspermia in a Milky Way-like Galaxy}

\author[0000-0003-0121-6113]{Raphael Gobat}
\affiliation{Instituto de F\'{i}sica, Pontificia Universidad Cat\'{o}lica de Valpara\'{i}so, Casilla 4059, Valpara\'{i}so, Chile}

\author[0000-0003-4923-8485]{Sungwook E. Hong \begin{CJK*}{UTF8}{mj}(홍성욱)\end{CJK*}}
\affiliation{Korea Astronomy and Space Science Institute, 776 Daedeok-daero, Yuseong-gu, Daejeon 34055, Republic of Korea}
\affiliation{Natural Science Research Institute, University of Seoul, 163 Seoulsiripdaero, Dongdaemun-gu, Seoul, 02504, Republic of Korea}

\author{Owain Snaith}
\affiliation{GEPI, Observatoire de Paris, PSL Research University, CNRS, Place Jules Janssen, 92190, Meudon, France}

\author[0000-0001-9991-8222]{Sungryong Hong \begin{CJK*}{UTF8}{mj}(홍성용)\end{CJK*}}
\affiliation{Korea Astronomy and Space Science Institute, 776 Daedeok-daero, Yuseong-gu, Daejeon 34055, Republic of Korea}

\correspondingauthor{Sungwook E. Hong}
\email{swhong@kasi.re.kr}

\begin{abstract}
We study the process of panspermia in Milky Way-like galaxies by modeling the probability of successful travel of organic compounds between stars harboring potentially habitable planets. To this end, we apply the modified habitability recipe of \citet{Gobat2016} to a model galaxy from the MUGS suite of zoom-in cosmological simulations. 
We find that, unlike habitability, which only occupies narrow dynamic range over the entire galaxy, the panspermia probability can vary be orders of magnitude between the inner ($R, b = 1-4\kpc$) and outer disk. However, only a small fraction of star particles have very large values of panspermia probability and, consequently, the fraction of star particles where the panspermia process is more effective than prebiotic evolution is much lower than from na\"ive expectations based on the ratio between panspermia probability and natural habitability.
\end{abstract}


\section{Introduction}\label{sec:intro}

The notion that living organisms could travel between celestial bodies is almost as old as the concept of habitable worlds (e.g., Lucian's \emph{Vera Historia}, Kepler's \emph{Somnium}, or Voltaire's \emph{Microm\'{e}gas}). Already present in embryonic form in some ancient mythologies, it was first formally named \emph{panspermia} 25 centuries ago by the philosopher Anaxagoras.
The development of microbiology in the 19$^{\text{th}}$~century opened up the possibility that such passage, rather than being directed, could take the form of the accidental propagation of simple \emph{seeds} of life \citep[e.g.,][]{Arrhenius1908}. 
This idea eventually spread through popular culture and became a minor staple of early 20$^{\text{th}}$ century speculative fiction, often playing on fears of invasion and contamination. 

In the modern era of astrophysics, the concept of panspermia was famously embraced and developed by F.~Hoyle and C.~Wickramasinghe after the discovery of organic compounds in the interstellar medium \citep[e.g.,][]{HoyleWickramasinghe1977,HoyleWickamasinghe1978,Hoyle1983,Hoyle1986}. 
Although never gaining widespread acceptance, likely due to a combination of practical and ideological factors (i.e., long odds and a perceived non-necessity in explaining our world), panspermia has been the subject of a steady number of studies since then. 
It has experienced a resurgence lately with the discovery of multiple, possibly habitable exoplanetary systems \citep{Gillon2016,Zechmeister2019} and the recent crossing through the solar system of hyperbolic trajectory comets of probable interstellar origin \citep{Meech2017,HiguchiKokubo2019}, starkly illustrating the possibility of matter exchanges between unbound star systems. 

Modern studies involving panspermia broadly fall into three categories.
The first concerns practical evaluations of the survivability of micro-organisms to the various potentially lethal events that panspermia involves. 
Namely, their ejection from and re-entry onto planetary surfaces \citep[e.g.,][]{Melosh1988,Burchell2001,Mastrapa2001,Stoffler2007,Burchell2007,Price2013,Pasini2015} and transit through the harsh radiation environment of interplanetary and interstellar space \citep[e.g.,][]{Weber1985,Horneck1994,Secker1996,Horneck2001,Wickramasinghe2003,Yamagishi2018}. 
These show that bacterial spores can survive hypervelocity impacts, as well as prolonged exposure to a combination of hard vacuum, low temperatures, and ionizing radiation. 
Although experiments were understandably not carried out over the timescales expected for interstellar panspermia, they nevertheless suggest that a small but non-insignificant fraction of spores could survive the kilo- or mega-years of transit, especially when embedded in even a thin mantle of carbonaceous material.

The second type of study tries to estimate the timescale and types of mass transfer between planets orbiting a common host star, typically by the exchange of meteoroids \citep[also called \emph{litho-panspermia};][]{Wells2003,Gladman2005,Krijt2017,Lingam2017}, or between stellar systems \citep[e.g.,][]{Melosh2003,Lingam2018}, with the radiation pressure on small grains \citep[also called \emph{radio-panspermia};][]{Wallis2004,Napier2004,Wesson2010,Lingam2021}. 
To these, we can also include speculations on the intentional seeding of other stellar systems via technological means \citep[or \emph{directed panspermia}; e.g.,][]{CrickOrgel1973}.

The last category concerns statistical investigations of panspermic propagation through stellar systems or galaxies using analytical or simple numerical models \citep{Adams2005,Lin2015,Lingam2016,Ginsburg2018,Dosovic2019,CarrollNellenback2019}, or of the effect of a spatially variable probability of habitable planets \citep[\emph{habitability}; e.g.,][]{Gonzalez2001,Lineweaver2004,Gowanlock2011} on the viability of panspermia. 

Here we couple the products of a hydrodynamical simulation of a Milky Way-like galaxy \citep{Stinson2010,Nickerson2013} with a modified galactic habitability model \citep[][hereafter GH16]{Gobat2016} to investigate how the probability and efficiency of panspermia vary with galactic environment. 
This paper is structured as follows: Sect.~\ref{sec:sim} and \ref{sec:hab} describe, respectively, the numerical simulation and habitability model that constitute the base of this study, while Sect.~\ref{sec:pans} presents the mathematical formalism we use for computing the probability of panspermia. 
We describe and discuss our results in Sect.~\ref{sec:res} and summarize our conclusions in Sect.~\ref{sec:concl}.

\section{Simulation Data}\label{sec:sim}

The \emph{McMaster Unbiased Galaxy Simulations} (MUGS) is a set of 16 simulated galaxies carried out by \citet{Stinson2010} and \citet{Nickerson2013}. These simulations made use of the cosmological zoom method, which seeks to focus computational effort into a region of interest, while maintaining enough of the surrounding large-scale structure to produce a realistic assembly history. To accomplish this, the simulation was first carried out at low resolution using N-body physics only. Dark matter halos were then identified, and a sample of interesting objects selected. The particles making up, and surrounding, these halos were then traced back to their origin, and the simulation carried out again with the region of interest simulated at higher resolution. The sample of galaxies was selected to have a minimal selection bias on the merging history or spin parameter. To be eligible for re-simulation halos had to have a mass between  $5\times 10^{11}$ and $2\times 10^{12}\,{\rm M}_\odot$, and be isolated for any object with a mass greater than $5\times 10^{11}$ by 2.7 Mpc. MUGS is therefore able to reproduce realistic infall and merging histories. Furthermore, they are able to reproduce the metallicity gradients seen in observed galaxies \citep[e.g.,][]{Pilkington2012, Snaith2016}, as well as features such as discs, halos and bulges. 

The MUGS galaxies were simulated in the context of a $\Lambda$CDM cosmology in concordance with the \emph{Wilkinson Microwave Anisotropy Probe} (WMAP) 3-year result \citep{Spergel2007},  with $(h, \Omega_m, \Omega_\Lambda, \Omega_b, \sigma_8)=(0.73,0.24, 0.76, 0.04, 0.79)$. The simulations were carried out down to $z=0$ using the \texttt{GASOLINE} \citep{Wadsley2004} smoothed-particle hydrodynamics code (SPH). 

In the high-resolution region a gravitational softening length of 310 pc was used, with a hydrodynamical smoothing length of 0.01 times the softening length. The masses of the dark matter, gas and stellar particles were $1.1\times 10^6$, $1.1\times 10^6$, and $6.3\times 10^6\Msun$ respectively. In order to reproduce the baryonic properties of observed galaxies the simulations made use of ultraviolet (UV) background radiation, and metal-based cooling at low temperatures \citep{Shen2010}. These are based on results from \texttt{CLOUDY} \citep{Ferland1998}. To govern star formation, the simulations used a Schmidt-Kennicutt relation to relate the gas density to the star formation rate \citep[SFR;][]{Kennicutt1998}. This is further influenced by feedback processes from supernovae, which release energy and metals into the interstellar medium (ISM; see \citealt{Stinson2006} for details). A Kroupa Initial Mass Function  \citep[IMF;][hereafter KTG93]{Kroupa1993} was used to calculate the stellar yields produced by each star particle\footnote{It is important to remember that each star particle is an ensemble of stars with a range of masses but the same metallicity. This is a consequence of the resolution limits of simulations, which cannot follow individual stars through time.}. The slope of the IMF controls the ratio of high mass to low mass stars in a given simple stellar population (SSP), and each star particle is essentially an SSP. The effect of the fraction of high mass stars is that it directly effects the ratio of Oxygen to Iron, because Oxygen is produced by core-collapse supernovae, while Iron is predominantly produced by SNIa. Thus, the IMF affects the stellar yields from star particles\footnote{\citet{Pilkington2012} found that the metallicity, $Z$, is calculated as ${\rm O+Fe}$, meaning that it is 1.8 times lower then expected. This may have a small impact on the cooling rate of gas, etc.}. Metals are allowed to diffuse between gas particles, using the method outlined in \citet{Wadsley2008}, in order to improve mixing.

\begin{figure*}[bt]
\centering
\plotone{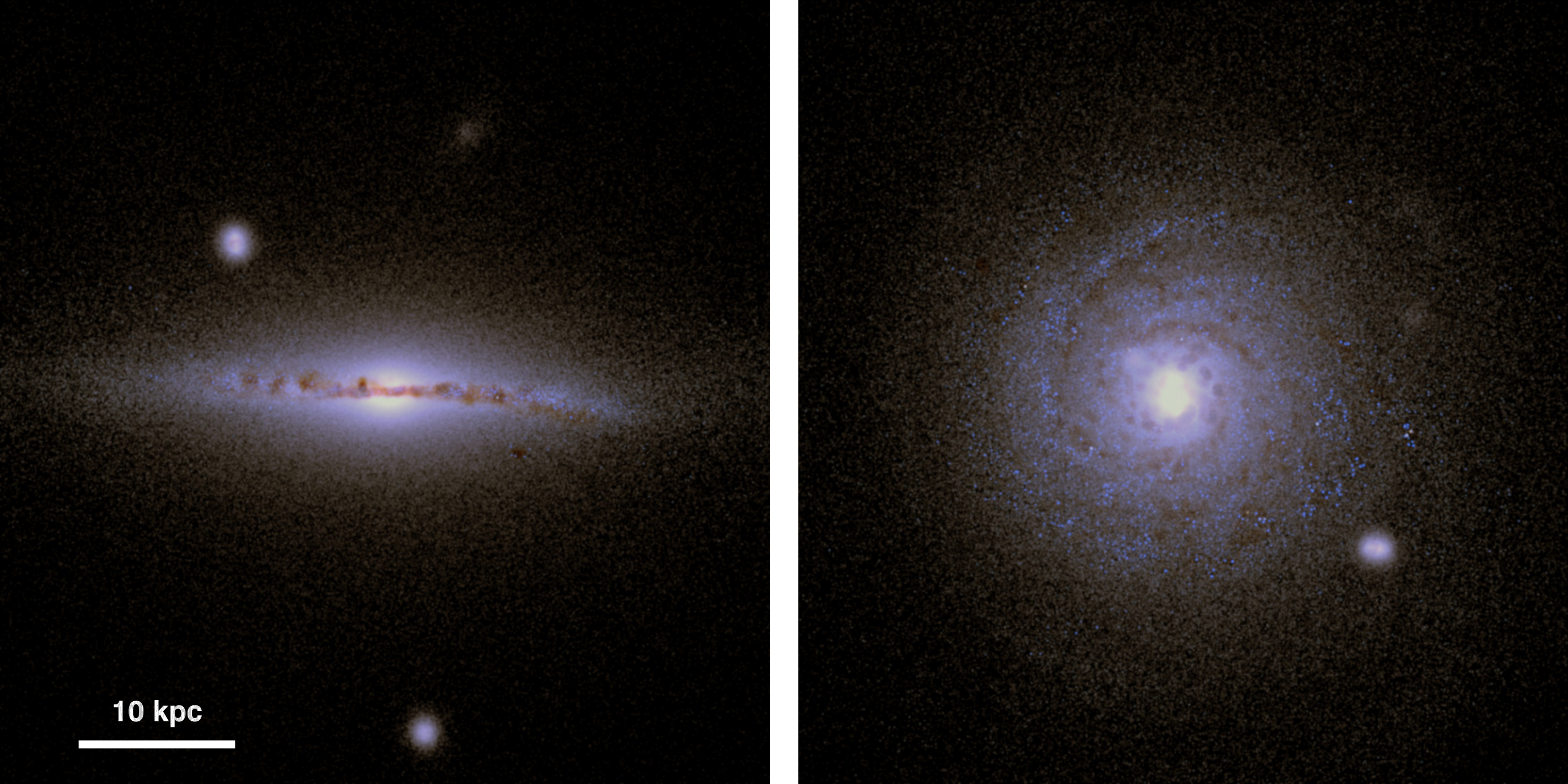}
\caption{Mock $UVJ$ color images of the simulated galaxy {\tt g15784} \citep{Stinson2010,Nickerson2013}, for both edge-on (\emph{left}) and face-on (\emph{right}) orientations, using star and gas particles, and assuming \citet{BC03} stellar population models and a simple dust attenuation model \citep{LiDraine2001} with a gas-to-dust ratio of 0.01 at solar metallicity. 
Additionally, we include line emission from star particles with ages $\leq 50\Myr$, following \emph{case B} recombination \citep{Osterbrock2006} and metallicity-dependent line ratios \citep{Anders2003}. 
All panels are $50\kpc$ across and have a resolution of $100\pc$. Two spheroidal satellites can be seen above and below the galactic plane, respectively.}
\label{fig:g15784}
\end{figure*}

We select one galaxy from MUGS for this study, {\tt g15784}, which is characterized by a quiescent merger history. The galaxy has a total and stellar mass of $1.5 \times 10^{12}$ and $1.14 \times 10^{11} \Msun$, respectively, and is therefore slightly larger than the Milky Way \citep{Licquia2015}. {\tt g15784} has not experienced a merger with an object more massive than a 1:3 ratio since $z_{\rm lmm} \simeq 3.42$, similar to our own Galaxy \citep[e.g.,][]{Stewart2008,Kruijssen2019, Naidu2021}. Several spheroidal galaxies can be seen to orbit {\tt g15784} within $25\kpc$ of the galactic center (see Fig.~\ref{fig:g15784})\footnote{For more details see \url{https://mugs.mcmaster.ca/g15784.html}}. For an overview of the detailed chemical properties of this galaxy, see \citet{Gibson2013} and \citet{Snaith2016}, among other works.

\begin{figure}[bt]

\includegraphics[width=0.495\textwidth]{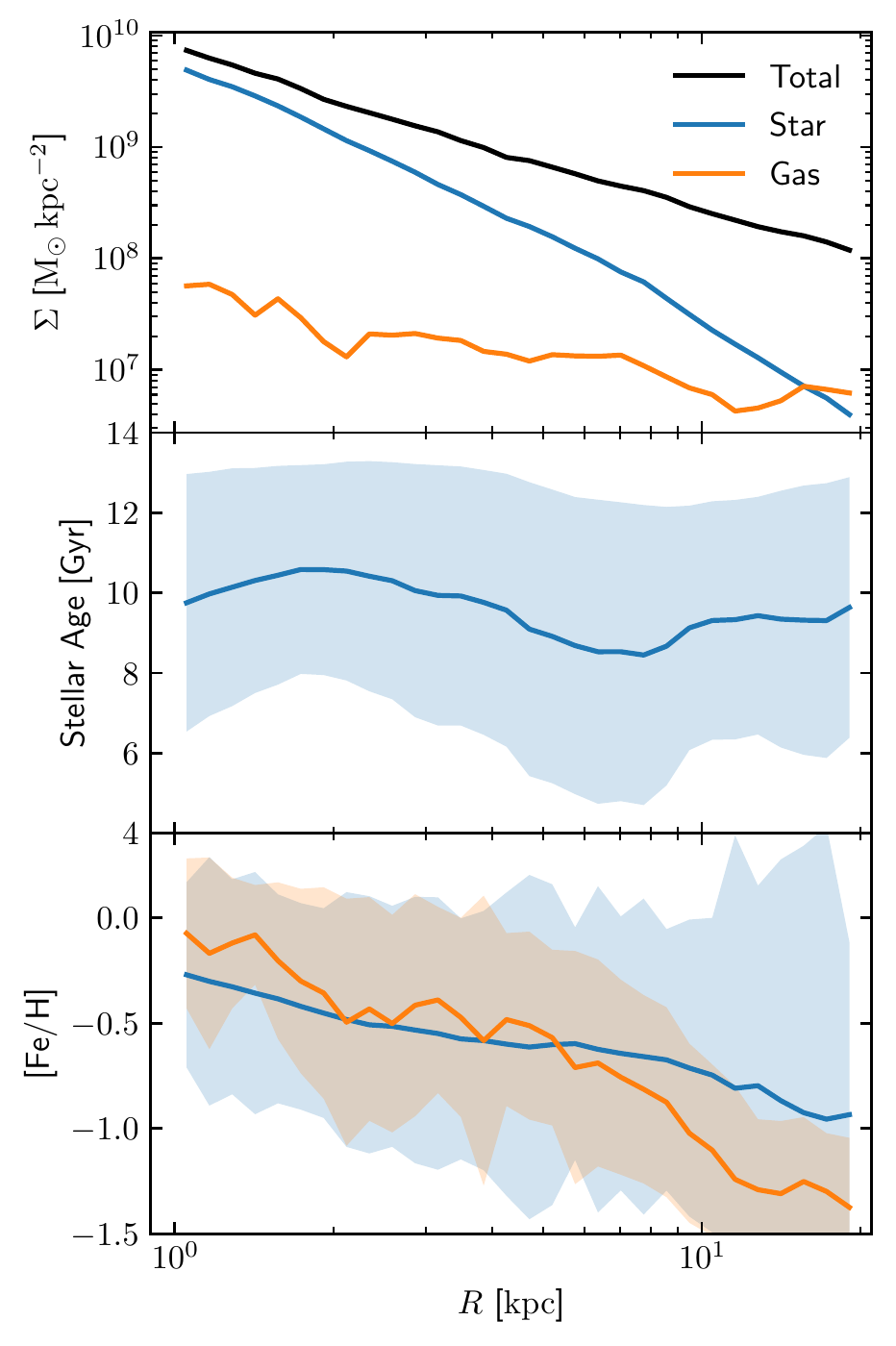}

\caption{Face-on radial profiles of the surface density (\emph{top}), stellar age (\emph{middle}), and the metallicity (\emph{bottom}) of {\tt g15784} at $z = 0$.
Shaded regions show the standard deviation within a given radial bin.}
\label{fig:g15784_profile}
\end{figure}

Fig.~\ref{fig:g15784_profile} shows the face-on radial profiles of the surface density, stellar age, and the metallicity of {\tt g15784}.
By taking the mean value of [Fe/H] in a series of radial bins between $3\kpc$ and $20\kpc$ we find this galaxy has a metallicity gradient of $-0.012\,{\rm dex/kpc}$. 
Outside of around $10\kpc$ the stars have a mean formation time of $6.5-7\Gyr$, rising from a formation time of $3.6\Gyr$ at a $1.5\kpc$. The SFR at $z=0$ is approximately $2.5 \Msun\,{\rm yr}^{-1}$.
One important caveat is that the bulge-to-disc light ratio ($\log ({\rm B/D})$) for this galaxy is larger than expected for the Milky Way, reaching a value of $-1.6$ for {\tt g15784} \citep{Brook2012}, compared to $+0.14$ for the Milky Way \citep{McMillan2011}. This is a known difference between the Milky Way and MUGS galaxies \citep{Stinson2010, Allen2006}.
This is the effect of insufficient feedback and resolution to capture all the physical processes required to produce a realistic simulated galaxy. 

From now on, we only consider the star particles located between $1 - 25\kpc$ from the mass-weighted center of the {\tt g15784}.
We ignore the inner 1$\kpc$ since it is not well resolved. We find a massive knot of particles in the inner few softening lengths, and the high stellar and gas density of this region might thus be an artifact of the simulation \citep{Stinson2010}. The volume we consider contains 977,304 star particles, which we separate into three distinct regions: (1) the {\tt CentralDisk}, located in galactocentric radius $R = 1 - 4 \kpc$ with a thickness of 1\,kpc along the galactic plane; (2) the {\tt DiskHalo}, which corresponds to both the inner disk ($R = 1 - 4 \kpc$ with larger thickness than {\tt CentralDisk}), outer disk ($R = 4 - 20\kpc$), and galactic halo outside the disk; and (3) the {\tt Spheroids} that include two orbiting spheroids.


\section{Levels of Habitability}\label{sec:hab}

For each star $6.3 \times 10^6 \Msun$ particle, we compute a measure of \emph{habitability}, that is, the fraction of main-sequence low-mass stars with terrestrial planets within their habitable zones (HZs). 
We mostly follow the model described in \citetalias{Gobat2016} and refer to that paper for more details on its various components. 
A quick summary of its formalism is as follows. 
The habitability fraction $f_{\rm hab}$ of a given star particle at the time $t = t_z$ is defined as
\begin{multline}\label{eq:hab}
f_{\rm hab} = \frac{1}{n_{\star}}\int_0^{t_z-t_{\rm min}} \d t \, (1-V_{\rm irr}) \Psi \\
\times \int_{m_{\rm min}}^{1.5 \Msun} \d m \, m\phi\,\mathcal{H}(t_{\rm MS}-t) w_{\rm h} 
\end{multline}
with
\begin{equation}
n_{\star}(t_z) = \int_0^{t_z} \d t \, \Psi \int_{m_{\rm min}}^{m_{\rm max}} \d m \, m\phi\,\mathcal{H}(t_{\rm MS}-t)
\end{equation}
\begin{multline}
w_{\rm h}(m,Z,t) = f_{\rm T}\,\mathcal{H}(1.5\Msun-m) \\
\times \left(\int_{r_{\rm h,i}}^{r_{\rm h,o}} \d r \, P^{\beta_{\rm P}} \frac{\d P}{\d r} \right) \big/ \left(\int \d r \, N\frac{\d P}{\d r} \right)
\end{multline}

\begin{equation}
f_{\rm T}(Z) = \mathcal{H}(Z-Z_{\rm min}) \left[ f_{\rm T,0} - f_{\rm HJ} \left( \frac{Z}{Z_{\odot}} \right)^{\alpha_{\rm P}} \right] \, ,
\end{equation}
where $\mathcal{H}(x)$ is the Heaviside step function, and $(m_{\rm min}, m_{\rm max}) = (0.1 \Msun, 100 \Msun)$ are the minimum and maximum mass of a star, respectively.
$\Psi(t)$ is the star particle's star formation history (SFH)\footnote{Since star particles are created just once due to the construction of the numerical simulation, this simply corresponds to a delta burst, to which we assign a duration of $100\Myr$.}, $\phi(m)$ is its IMF, and $V_{\rm irr}(t)$ is its (time-dependent) volume fraction irradiated by SN. 
$t_{\rm MS}(m,Z)$ is the main-sequence lifetime of a star with mass $m$ and metallicity $Z$, and $r_{\rm h,i}(m,Z,t)$ and $r_{\rm h,o}(m,Z,t)$ are the inner and outer radii of its HZ, respectively.
$P(r,m)$ are the orbital periods of planets orbiting that star, whose distribution follows a simple power-law $\d N \propto P(r,m)^{\beta_{\rm p}} \d P$ \citep[e.g.,][]{Cumming2008,Petigura2013,Burke2015}.
$f_{\rm T}(Z)$ is the fraction of stars with terrestrial planets 
\citepalias[\emph{Case 2} of the metallicity dependence in][]{Gobat2016}, and $f_{\rm HJ}$ is the fraction of solar-metallicity stars with short-period gas giants (or, \emph{hot Jupiters}), with $Z_{\rm min}=0.1Z_{\odot}$ being a metallicity threshold below which we assume that terrestrial planets cannot form.
We do not consider the contribution of the active nucleus in the galactic core, since the distance at which its additional radiative input significantly alters HZs is somewhat smaller \citepalias{Gobat2016} than the radius of the simulation volume we ignore. 
The number of (potentially) habitable stellar systems in each star particle is therefore given as $n_{\rm hab}=f_{\rm hab} n_{\star}$.

The values of each parameter we used in this paper are the same as in \citetalias{Gobat2016}: $(f_{\rm HJ}, \alpha_{\rm P}, \beta_{\rm P}) = (0.012, 2, -0.7)$, with the following exceptions.
First, $t_{\rm min}$, the time after which a planet is considered able to host life, is decreased from $1\Gyr$ to $500\Myr$ in the \emph{natural} case (see below).
Also, we set $f_{\rm T,0}=1$ \citep[similar to][]{Zackrisson2016} to reconcile the predictions of the model with \emph{Kepler} constraints \citep{He2019}, which previously differed by a factor of $\sim 2$ \citepalias{Gobat2016}.
Finally, we modify the inner radius of the HZ \citep[Eq.~(3) in][]{Kopparapu2013}, to account for increased flare activity in low-mass stars. 
In essence, we increase the luminosity $L$ of every star by the energy-integrated cumulative distribution of flare frequency (which we approximate as a power-law with an index of $-0.8$), limited by its mass-dependent maximum flare energy, from \citet{Davenport2016}. Since we are only concerned with average quantities, we simply weigh this additional luminosity by the fraction of flaring stars to derive the fractional increase per stellar mass.
This has the effect of slightly reducing the habitability weight $w_{\rm h}(m,Z,t)$ of sub-solar mass stars. However, the effect is mild enough that it does not modify our results.
\footnote{In a similar way, atmospheric erosion by intense stellar winds might also push the inner edge of the HZ outwards without impacting its outer radius \citep[e.g., see][]{Dong2018}, effectively narrowing it further around M-dwarfs as they age.}

\begin{figure}
\centering
\includegraphics[width=0.495\textwidth]{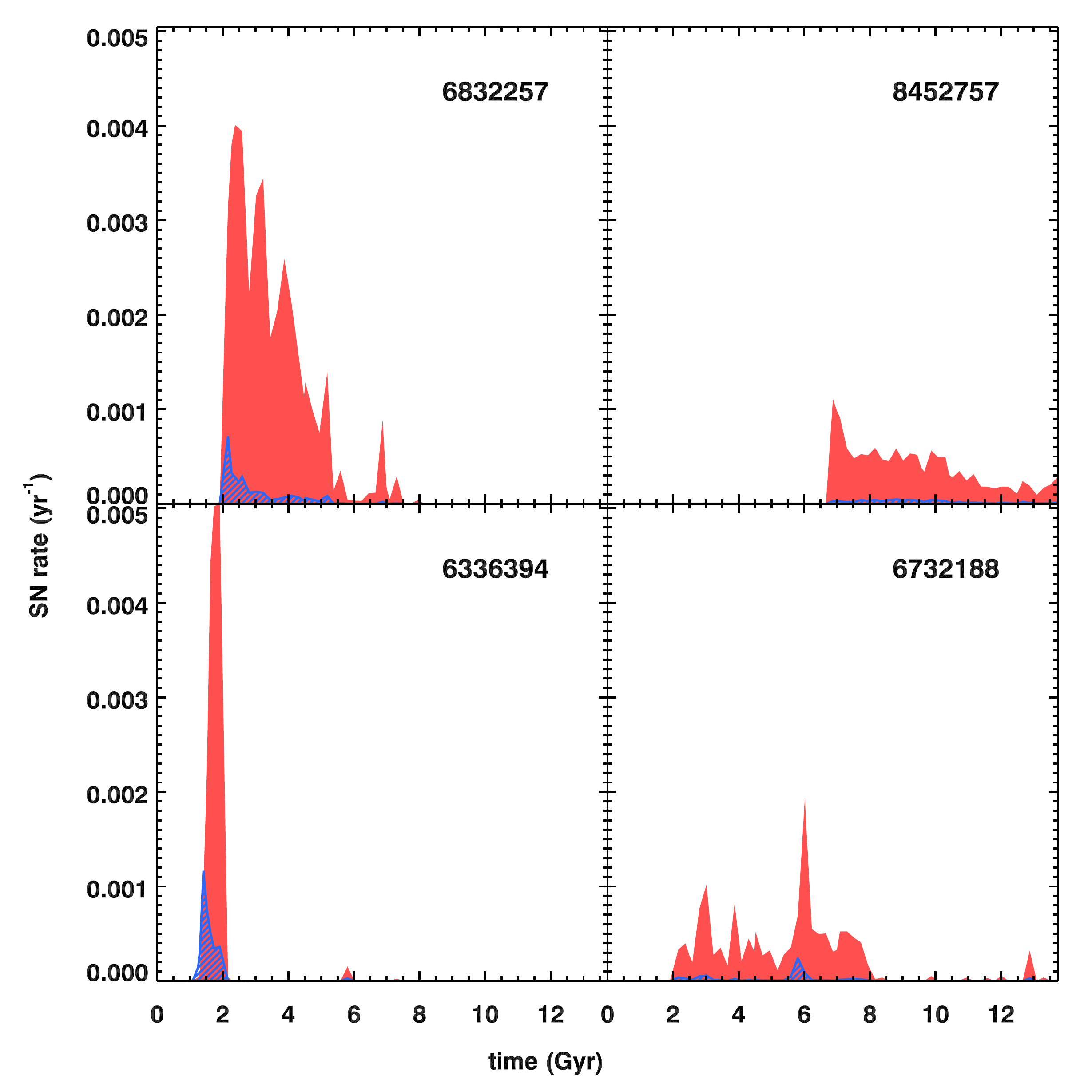}

\caption{Rates of Type Ia (red) and Type II (blue) supernovae for selected star particles in {\tt g15784}. 
Particle IDs are shown in the upper right corner of each panel.}
\label{fig:snr}
\end{figure}

We compute supernova rates (SNR) for each star particle based on its SFH and neighboring particles at each time step, using simulation snapshots from $z=10.8$~to $z=0$, spaced by $\sim$200\,Myr. 
We consider Type Ia and II supernovae, with
\begin{align}
{\rm SNR}_{\rm Ia}(t) &= 
\eta_{\rm WD} \int_{m_{\rm min}}^{8 \Msun} \d m \, m  \nonumber \\
& \, \times \int_{\tau_{\rm Ia}+t_{\rm MS}(8\Msun)}^t \!\!\!\!\!\! \d t' \, \Psi(\tau) \phi(m,\tau) \\
{\rm SNR}_{\rm II}(t) &= \Psi(t)\int_{8\Msun}^{m_{\rm max}} \d m \, m\,\phi(m,t) \, , \label{eq:SNrate}
\end{align}
where $\eta_{\rm WD}=0.01$ is the white dwarf (WD) conversion rate \citep{Pritchet2008}, $\tau=t'-t_{\text{MS}}(m)-\tau_{\text{Ia}}$, 
and $\tau_{\text{Ia}}=500\Myr$ is the average delay time between stellar death and the detonation of the WD \citep{Raskin2009}. 
An example of SNRs for selected star particles in {\tt g15784} is given in Fig.~\ref{fig:snr}. The volume irradiated by supernovae within star particle $i$ is then
\begin{multline}\label{eq:SNvolume}
V_{{\rm irr},i}(t) = \frac{\mathcal{H}(t_{\rm rec}-t)}{r_p^3} \\
\times \sum_j v_{ij} \left({\rm SNR}_{{\rm Ia},j} r_{\rm Ia}^3 + {\rm SNR}_{{\rm II},j} r_{\rm II}^3 \right) \, ,
\end{multline}
where $(r_{\rm Ia}, r_{\rm II})=(0.3 \pc, 0.5 \pc)$ are the lethal radii of Type Ia and II supernovae, respectively \citepalias{Gobat2016}. 
$v_{ij}$ is the fractional volume where the two $6.3 \times 10^6 \Msun$ particles overlap (with $v_{ii}=1$), computed assuming that each corresponds to a spherical volume of radius $r_p$, set by the gravitational softening length, homogeneously filled with stars.
As in \citetalias{Gobat2016}, the recovery time $t_{\rm rec}$ is set to an arbitrarily large value so that $V_{\rm irr}$ is always positive.

\begin{figure*}[p]
\centering
\includegraphics[height=0.3\textheight]{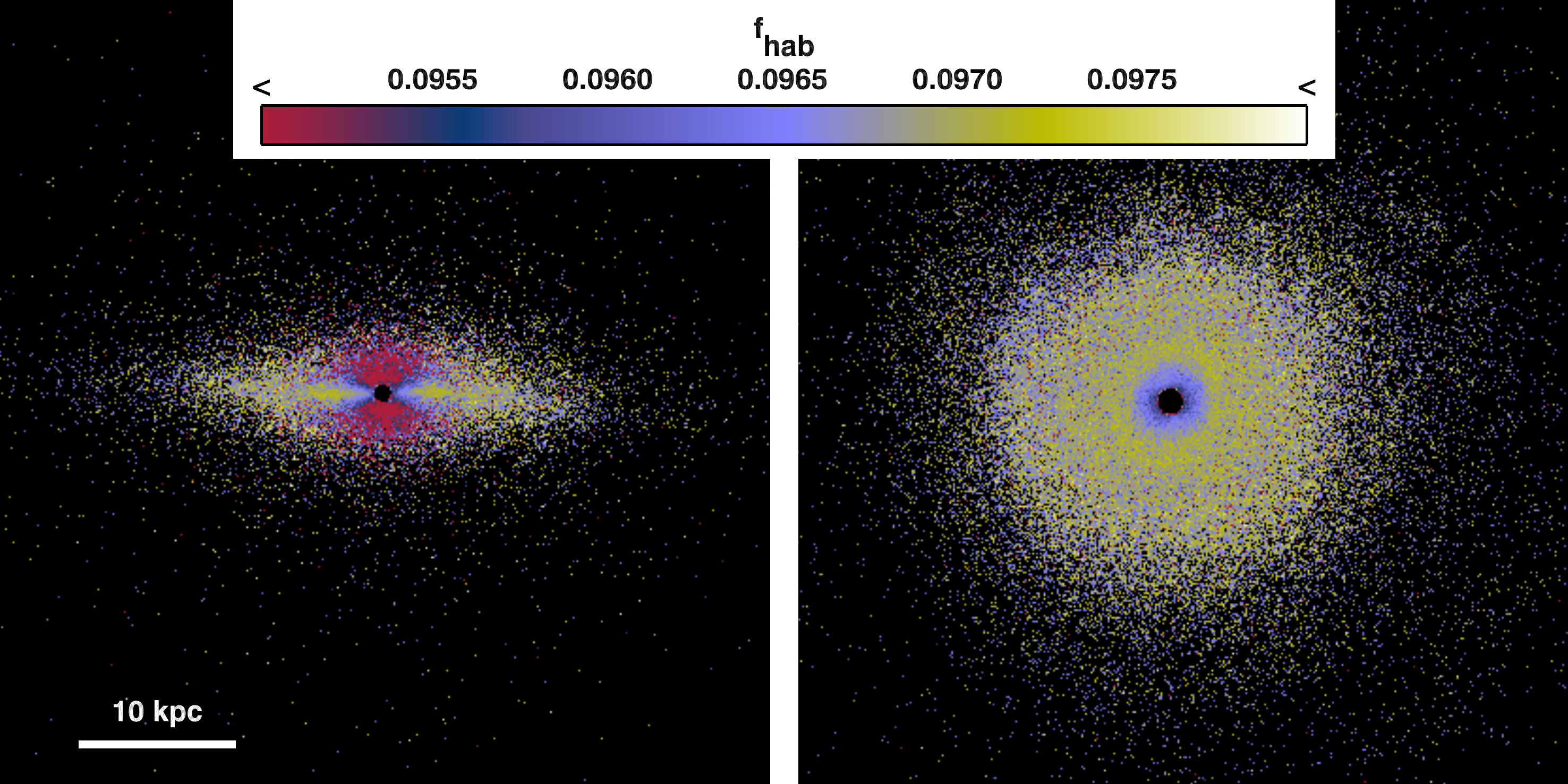}\\
\includegraphics[height=0.3\textheight]{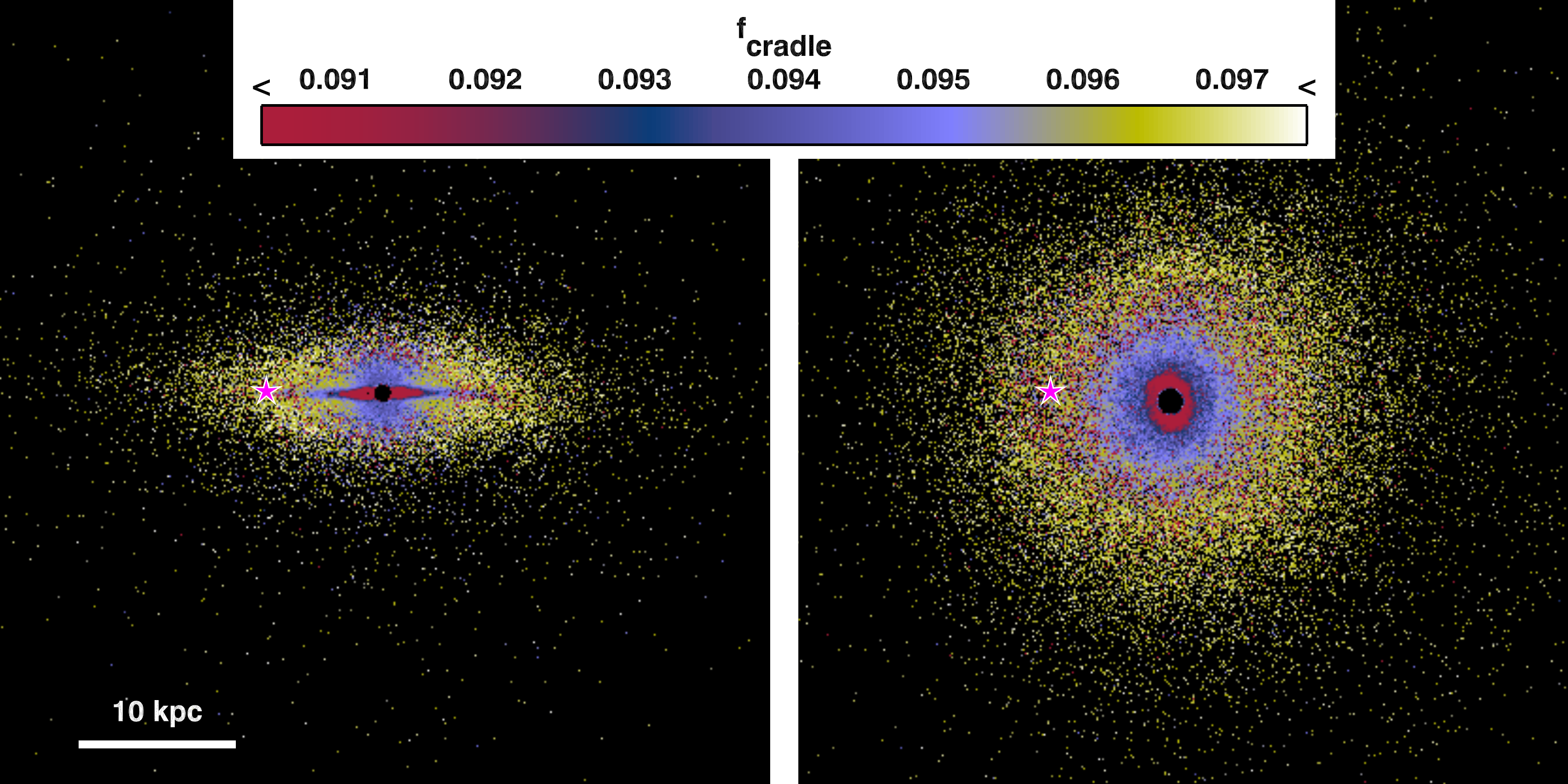}\\
\includegraphics[height=0.3\textheight]{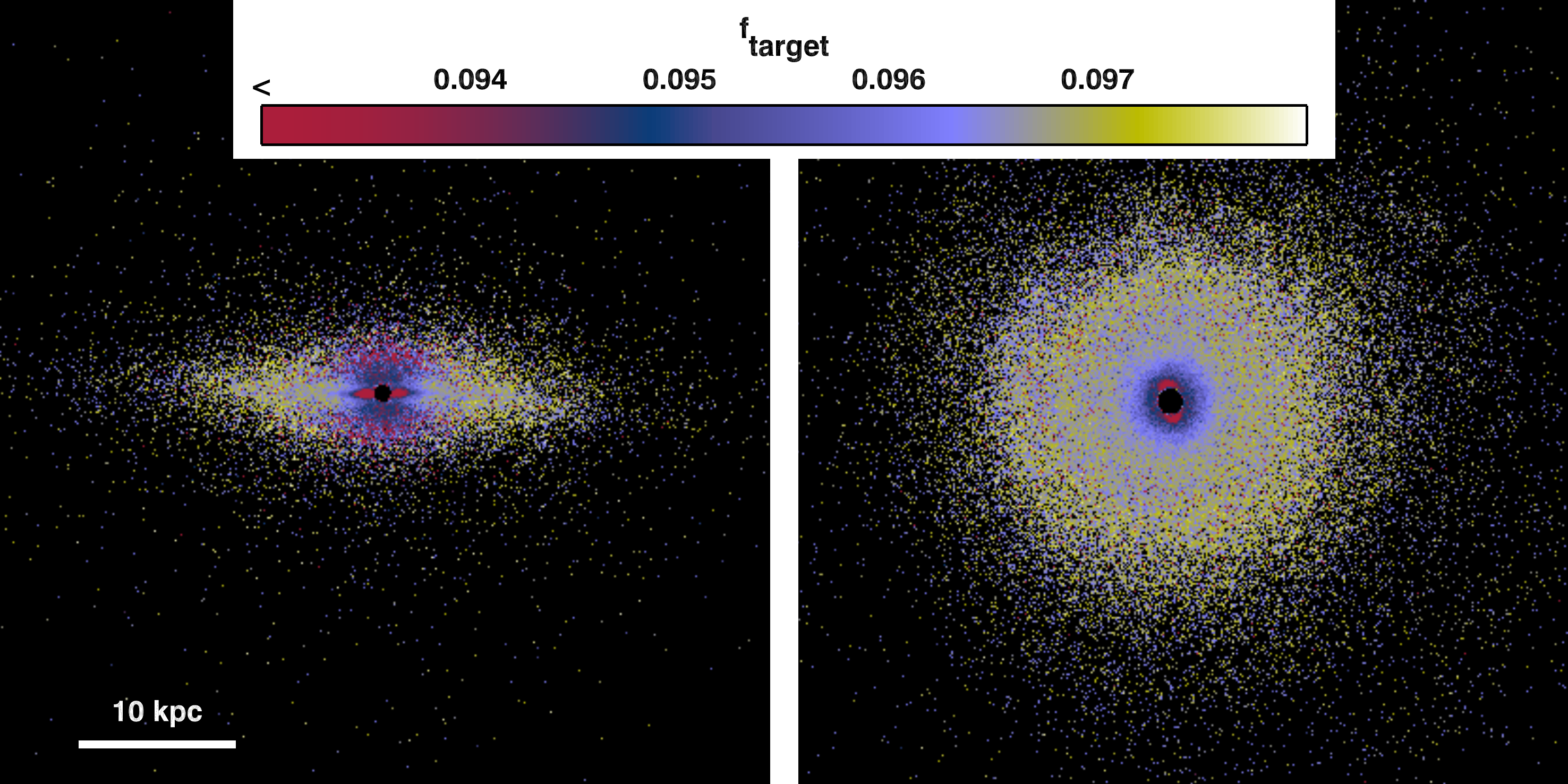}
\caption{
Similar to Fig.~\ref{fig:g15784}, except showing the median value of the \emph{natural} habitability ($f_{\rm hab}$; \emph{top}), the fraction of possible \emph{cradles} ($f_{\rm cradle}$; \emph{middle}), and the fraction of possible colonization \emph{targets} ($f_{\rm target}$; \emph{bottom}) within each projected column at $z=0$ and in a $1\kpc$-wide slice passing through the center of {\tt g15784}. The magenta star corresponds to the approximate position of the Sun, if it were the actual Milky Way.}
\label{fig:fhab}
\end{figure*}

\begin{figure}[bt]
\centering

\includegraphics[width=0.495\textwidth]{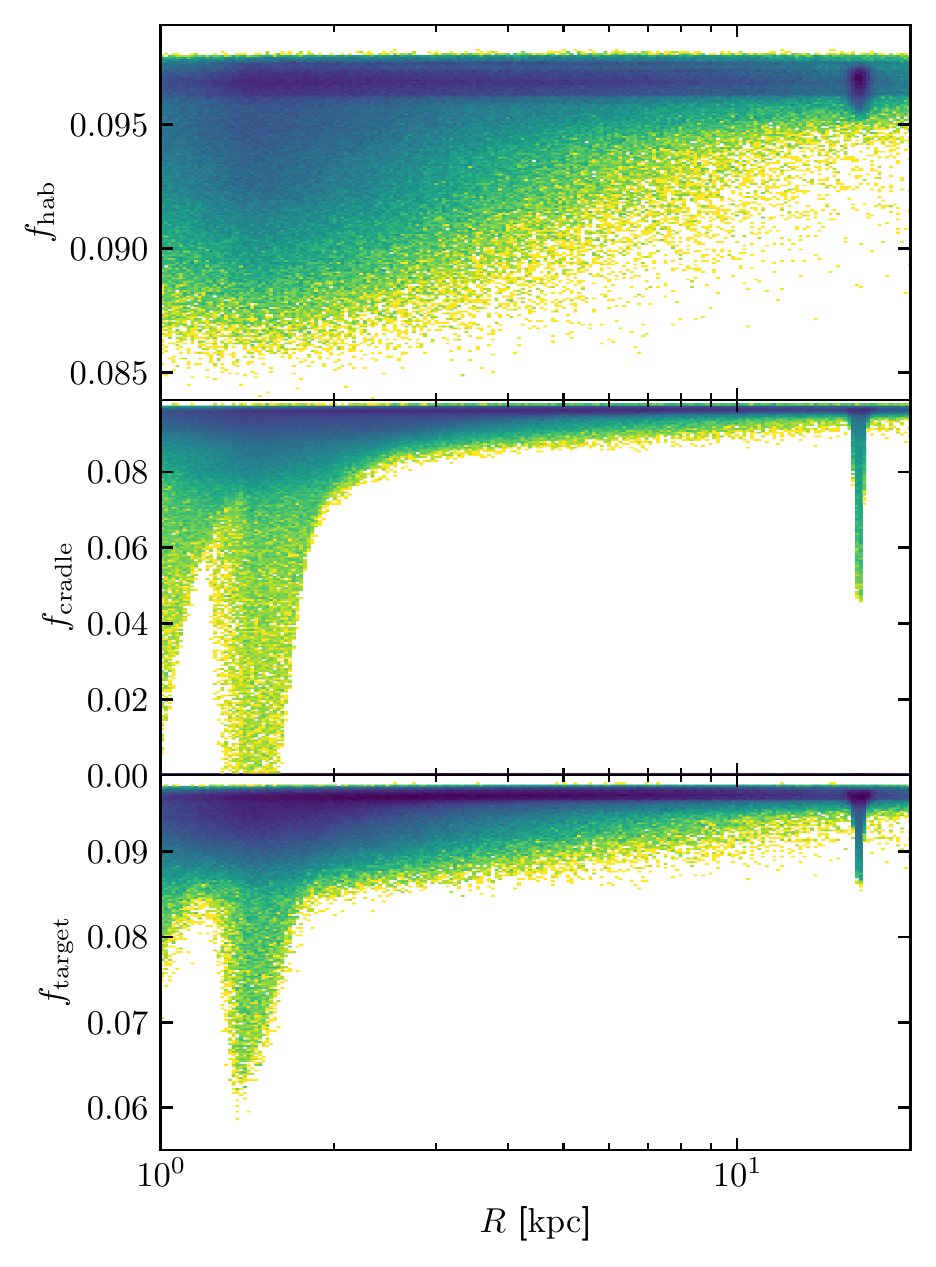}
\caption{The fraction of \emph{natural} habitable planets ($f_{\rm hab}$; \emph{top}), \emph{cradles} of potential civilizations ($f_{\rm cradle}$; \emph{middle}), and possible colonization \emph{targets} ($f_{\rm target}$; \emph{bottom}), as a function of galactocentric radius in {\tt g15784} at $z = 0$, drawn in a logarithmic scale.}
\label{fig:habrad}
\end{figure}

In a majority of cases, most of the SNR felt by individual star particles is contributed by neighboring particles, whose volumes overlap each other. This, together with their higher metallicity (to which $f_{\rm T}(Z)$ is anti-correlated), reduces the habitability of star particles within the crowded galactic bulge, as shown in Fig.~\ref{fig:fhab}. 
On the other hand, the higher density and SNRs found in spiral arms have little effect, although a smattering of low-habitability particles exists in these regions. 
The average $f_{\rm hab}$ of the disk reaches a (weak) maximum at $\sim 4-5 \kpc$ from the center (Fig.~\ref{fig:habrad}), decreasing smoothly further out due to an increased fraction of very low metallicity $f_{\rm hab}=0$ star particles. 
This behavior is somewhat similar to the model of \citet{Lineweaver2004} but in contrast with the radially monotonous trend from \citet{Gowanlock2011}. 
The galactic halo similarly contains a mix of lower and higher-habitability star particles, with a significant fraction of $< 0.1 Z_\odot$ non-habitable star particles and a slightly lower average $f_{\rm hab}$ than the disk.

Overall, we find that the range of positive $f_{\rm hab}$ is somewhat narrow at $z=0$, with most particles having values within $\sim 5\%$ of each other, which is similar to \citet{Prantzos2008}. This is not entirely surprising since the probability of terrestrial planets $f_{\rm T}$ is, in this model, only weakly correlated with metallicity and the effect of supernovae on habitability is not strong (\citetalias{Gobat2016}). However, the variation increases by order of magnitude at $z>0.5$, due to a significantly higher number of star particles below the $0.1 Z_{\odot}$ threshold (especially at large galactocentric radii) and a much higher SNR (especially in the core).

Finally, we consider two other levels of habitability in addition to the \emph{natural} criterion described above, where a planet is deemed habitable quickly and SN effects are small. These are intended to describe better the fraction of planets where life could arise or take hold (see Sect.~\ref{sec:pans_oc}), as opposed to the larger population with just benign surface conditions.
The first habitability level is meant to mirror the present state of the Earth, where we set $t_{\rm min}=4.5\Gyr$ and require that no SNs happened within $r_{\rm Ia}=r_{\rm II}=8\pc$ of planets in the last $t_{\rm rec}=50\Myr$. 
We regard this habitability level, $f_{\rm cradle}$, as describing the \emph{cradles} of potential civilizations, that is, the origin point of directed panspermia.
Also, we consider a less restrictive case, with $t_{\rm min}=1\Gyr$ and $t_{\rm rec}=1\Myr$, to describe the potential \emph{targets} of such directed panspermia ($f_{\rm target}$).
The requirement for a small but non-zero $t_{\rm rec}$ also allows us to sidestep the issue of transit of organisms through the radiation environment of supernova remnants.
Contrarily to the natural case, here the low-habitability star particles are concentrated within the {\tt CentralDisk}
, up to $\sim 4\kpc$ as shown in Figs.~\ref{fig:fhab} and \ref{fig:habrad} 
, due to the higher SN efficiency at suppressing habitability. 
The outer galactic disk also has lower averages of $f_{\rm cradle}$ and $f_{\rm target}$, due to the higher prevalence of recently-formed star particles, and the distributions of cradles and targets are similar.


\section{Modelling The Probability Of Panspermia}\label{sec:pans}

The probability that seeds of life departed from star particle $i$ at position $\vec{x}_i$ at time $t_i$ (hereafter, $(\vec{x}_i, t_i)$) to be successfully transplanted to star particle $j$ at $(\vec{x}_j, t_j)$ can be written as
\begin{multline}
\mathcal{F}(\vec{x}_i,t_i; \vec{x}_j, t_j) \\
= f_{\rm oc}(\vec{x}_i,t_i) f_{\rm target}(\vec{x}_j,t_j) f_{\rm esc}(\vec{x}_i) f_{\rm cap}(\vec{x}_j) \\
\times \int_{\mathbb{V}_i} \!\! \d\vec{x}'_i \int_{\mathbb{V}_j} \!\! \d\vec{x}'_j  w_{\rm damage} (\vec{x}'_j,t_j| \vec{x}'_i, t_i) \, .
\label{eq:prob_onebyone}
\end{multline}
where the definitions of $f_{\rm oc}$, $f_{\rm target}$, $f_{\rm esc}$, $f_{\rm cap}$, and $w_{\rm damage}$ are given progressively in the following sections. We note that, while we here apply Eq.~(\ref{eq:prob_onebyone}) to what are effectively groupings of millions of stars, it can also be used to describe panspermia between two individual stars.
Then the total amount transplanted to star particle at $(\vec{x}_j, t_j)$ is proportional to
\begin{equation}
f_{\rm pp}(\vec{x}_j, t_j) = \sum_{\vec{x}_i} \int_{0}^{t_j} \!\! \d t_i \, \mathcal{F}(\vec{x}_i,t_i; \vec{x}_j, t_j) \, ,
\label{eq:prob_onebyone2}
\end{equation}
We hereafter call $\mathcal{F}$ and $f_{\rm pp}$ the \emph{panspermia contribution} and \emph{panspermia probability}, respectively.\footnote{The actual probability that the panspermia process is successful may not be simply additive as assumed in Eq.~(\ref{eq:prob_onebyone2}), but this is beyond the scope of this paper.}
While $t_j$ should always be greater than $t_i$, $\vec{x}_i$ and $\vec{x}_j$ can be the same since a single star particle in our analysis contains millions of stars.
\deleted{In the following subsections, we will discuss each term that contributes to the panspermia probability.}

\subsection{Probability of Hosting Life}\label{sec:pans_oc}

First, $f_{\rm oc}(\vec{x}_i, t_i)$ is the probability that the star particle $i$ has planets bearing living organisms, or their complex organic precursors, at $t_i$.
For a star particle having these seeds, it should be habitable for a period sufficient for successful prebiotic evolution. This can be written as
\begin{equation}
f_{\rm oc}(\vec{x},t) = \int_{t-t_{\rm chem}}^t \!\!\!\! \d t' f_{\rm hab}(\vec{x},t') W_{\rm chem}(t-t') \, ,
\end{equation}
where $t_{\rm chem} \sim \mathcal{O}(10^{-1} \Gyr)$ is the typical timescale for prebiotic evolution \citep{mojzsis1996,vanzuilen2002,dodd2017}, and $W_{\rm chem}$ is a certain time-domain kernel between $f_{\rm hab}$ and $f_{\rm oc}$, respectively.
In practice, as $W_{\rm chem}$ is unknown and $t_{\rm chem}$ is much smaller than timescales used in this paper (e.g., timescales for star formation, local SN rates, and delay time for $f_{\rm hab}$). Therefore, for simplicity, we assume that the probability of hosting life is similar to the fraction of possible \emph{cradles} of life, that is, $f_{\rm oc}(\vec{x},t) \propto f_{\rm cradle}(\vec{x},t)$.

\subsection{Escape Fraction} \label{sec:pans_fesc}

$f_{\rm esc}(\vec{x}_i)$ is a weight proportional to the escape fraction of life spores, from habitable planets in the star particle $i$ to the ISM by natural processes.
In order to estimate $f_{\rm esc}$ in the case of natural panspermia, we assume that material carrying spores can be ejected from the surface of habitable terrestrial planets by surface impacts. The velocity of ejecta is \citep{Housen2011}
\begin{equation}\label{eq:vej}
v_{\rm ej}=u\cdot C_1 \left[ \frac{x}{a_e} \left( \frac{\rho}{\delta}\right)^{\nu} 
\right]^{-\frac{1}{\mu}} \left( 1-\frac{x}{n_2 R} \right)^{p_e}
\end{equation}
at a distance $x$ from the impact center, with $n_1 a\leq x\leq n_2 R$, where $R$ is the radius of the impact crater, and $u$ the impactor's velocity, which we assume to be equal to the average orbital velocity $v_{\text{orb,HZ}}(m_{\star})$ within the star's habitable zone.\footnote{The typical velocity of impactors may actually be closer to half of the planet's orbital velocity \citep[e.g., see][]{Collins2005}, which would naturally decrease the number of fragments reaching escape velocity. However, this would not change the ``normalized'' value of the escape weight $f_{\rm esc}$ which we use here.} 
For Eq.~(\ref{eq:vej}), we use parameter values for low-porosity rock given in \citet{Housen2011}, namely, $\mu=0.55$, $\nu=0.4$, $C_1=1.5$, $n_1=1.2$, $n_2=1$, $p_e=0.5$, $\rho/\delta\sim1$, and $a_e=0.0016$. 
The radius $R$ of the crater is related to the impactor's energy through 
\begin{equation}\label{eq:crater}
\frac{R}{{\rm km}} \sim \frac{1}{2}\left(\frac{E_{\rm kin}}
{9.1\times10^{24}~{\rm ergs}}\right)^{1/2.59}
\end{equation}
\citep{Hughes2003}, where $E_{\text{kin}}=\frac{1}{2} m_{\text{i}} v_{\rm o,HZ}^2$, with $m_{\text{i}}$~being the mass of the impactor. 
Assuming that the latter follows a distribution of the type $\propto m^{-1.6}$ \citep{Simon2017}, we can then compute a velocity distribution of ejecta (see Fig.~\ref{fig:ejecta1}).

Only a small fraction of the ejected material will have enough velocity to escape the planet and thus only an infinitesimal quantity of this material can ever reach stellar escape velocity in the simulation. 
On the other hand, we can expect that a fraction $w_{\rm esc}$ of material ejected into circumstellar orbit can be accelerated above stellar escape velocity by either gravitational interactions \citep[in the case of macroscopic fragments, which corresponds to lithopanspermia;][]{Melosh1988} or radiation pressure \citep[in the case of microscopic fragments;][]{Arrhenius1908}. 
Under the assumption that the host star mass does not determine the magnitude and frequency of these boosts, $w_{\rm esc}$ is only a function of orbital and stellar escape velocities within the HZ: 
\begin{equation}\label{eq:escphi}
w_{\rm esc}(m) = C\int_{r_{\rm h,i}(m)}^{r_{\rm h,o}(m)}\int_{v_{\rm esc}(r)}\phi_{\rm ej}(v)\d v\d r \, ,
\end{equation}
where $\phi_{\rm ej}$ is the quantity of ejected material as a function of velocity (see Fig.~\ref{fig:ejecta1})and $C$ is a normalizing constant.
Following this scheme, $w_{\rm esc}$ would be higher for higher-mass stars (Fig.~\ref{fig:ejecta1}), as the radius of the HZ increases with $L^{1/2}$ (that is, with $\sim M^{1.75}$). The same applies to lower-metallicity stars, which are brighter for the same mass. Since the magnitude and frequency of the boosts are unknown, a quantitative estimate for $w_{\rm esc}$ is however not possible. Instead, we compute relative quantities, rescaled to an arbitrary range of $\left[0,1\right]$ for our star particles at $z=0$. Integrating this over the age-truncated \citetalias{Kroupa1993} IMF then yields a weight 
\begin{equation}\label{eq:escweight}
f_{\rm esc}(t)=\int \d \, \log m \, \phi(m,t) w_{\rm esc}(m)N_{\rm impact}(t)
\end{equation}
for each star particle. Here we assume for simplicity that $w_{\rm esc}$ does not vary with the age of the stellar particle. However, we note that the frequency of impacts on Earth has decreased exponentially since the formation of the Solar System, which would result in a time-varying $f_{\rm esc}$. Since $w_{\rm esc}$ depends on the position of the HZ, the spatial distribution of $f_{\rm esc}$ within {\tt g15784} is similar to $f_{\rm hab}$ (Figs.~\ref{fig:fhab} \& 
\ref{fig:habrad}), being low within the {\tt CentralDisk}, reaching a maximum at $\sim 5\kpc$, and monotonously decreasing at larger galactocentric radii. For a more detailed description, see Appendix~\ref{appendix:wesc}.

\subsection{The Capture Fraction}\label{sec:pans_cap}

$f_{\rm cap}(\vec{x}_j)$ is the capture fraction of interstellar spores by target planets in the star particle $j$ by gravity. Here, we assume that stars are evenly distributed in each star particle, and that each star is separated from the others by a distance much greater than the size of its planetary system. 

The probability of interstellar objects captured by solar systems can be sensitive to the structure of planetary systems \citep[e.g.,][]{Ginsburg2018}. However, this fine structure lays considerably beyond the resolution of the simulation used here. Consequently, and for convenience, we assume $f_{\rm cap}(\vec{x}_j)$ to be simply constant on average.

\subsection{Traveling Seeds}\label{sec:pans_damage}

The last term of Eq.~(\ref{eq:prob_onebyone}) corresponds to the probability that spores originating from a habitable planet within the star particle $i$ ($\mathbb{V}_i$) at $t_i$ reach a target planet within the star particle $j$ ($\mathbb{V}_j$) at $t_j$.
Mathematically, it is calculated as the double volume integral of $w_{\rm damage}(\vec{x}'_j,t_j|\vec{x}'_i,t_i)$, the survival probability of spores in an interstellar object that starts at $(\vec{x}'_i, t_i)$ and arrives $(\vec{x}'_j, t_j)$ (hereafter, the \emph{damage weight}), where  $\vec{x}'_{i,j} \in \mathbb{V}_{i,j}$.
To compute this numerically, one needs to understand the spatial distribution of stars in each star particle, as well as consider the traveling distance ($\ell_{\rm surv}$), which is a function of the survival timescale of spores in the ISM ($t_{\rm surv}$), and how $w_{\rm damage}$ depends on $\ell_{\rm surv}$.

As mentioned in Sect.~\ref{sec:sim}, each star particle in {\tt g15784} is an ensemble of millions of stars whose total mass is about $6.3 \times 10^6 \Msun$.
While the volume of such an ensemble can be approximated by the cell from the Voronoi tessellation\footnote{Voronoi tessellation partitions a volume into multiple regions, based on a set of positions $\left\{ \vec{x}_i \right\}$. The Voronoi volume of a particle $\vec{x}_i$ is then defined as an ensemble of points whose closest particle is $\vec{x}_i$, that is, $\mathbb{V}_i \equiv \left\{ \vec{x} | |\vec{x},\vec{x}_i| < |\vec{x} - \vec{x}_j| \textrm{ for all } j \neq i \right\}$.} of star particles, the actual distribution of stars within the volume is unknown.
Here we assume that the stars in a star particle are uniformly distributed in its Voronoi volume by adopting such numerical limitation.
Furthermore, for a fast calculation of the panspermia probability, we simplify the geometry of the star particle $i$ to a sphere with center at $\vec{x}_i$ and radius $R_i \equiv \left( 3V_i / 4 \pi \right)^{1/3}$, where $V_i$ is its Voronoi volume.
Such assumptions will work well for small-volume star particles at a relatively dense region of {\tt g15784} because the stellar density gradient within the volume will be low, and the shape of the Voronoi cell will be close to the sphere due to numerous nearby cells.
On the other hand, our assumptions may poorly work for star particles at the outskirts of {\tt g15784}. However, these typically do not contribute much to our main results because of their low habitability and panspermia probability.

The survival time of microorganism spores under the pressures, temperatures, and radiation flux expected for panspermia is uncertain. 
All in-situ experiments so far have been done in Earth orbit \citep[e.g.,][]{Horneck2001,Onofri2012,Kawaguchi2013} and their conclusions are typically extrapolated to the requirements for in-system panspermia (e.g., between the Earth and Mars). 
On the other hand, they do not reflect the likely conditions of interstellar panspermia; in particular, the survival of spores under hard radiation has been found to increase at low temperatures \citep{Weber1985,Sarantopoulou2011}. 
Only a few biological studies concerning an interstellar transit have been carried out \citep{Weber1985,Koike1992,Secker1994}, which suggest that shielding by a rock or carbonaceous material would be required to ensure the survival of a sufficient number of spores.
With no clear constraints on the survival timescale, we assume a conservative choice of $t_{\rm surv} \sim \mathcal{O}(1\,{\rm Myr})$.
For simplicity, we also assume that $t_{\rm surv}$ is negligible compared to the evolution timescale of the Milky Way so that we can use the characteristics of star particles from the snapshot data at $z = 0$ for both $t_i$ and $t_j$.

The survival timescale $t_{\rm surv}$ can be rewritten in terms of the travel distance scale $\ell_{\rm surv} \equiv \bar{v}_{\rm oc} t_{\rm surv}$, where $\bar{v}_{\rm oc}$ is the mean velocity of the spores.
As already discussed in Sect.~\ref{sec:pans_cap}, the distance between different solar systems is much larger than the typical size of a solar system.
Therefore, we assume that, although not exactly zero, the probability that interstellar objects would significantly change their trajectories or speed due to the gravity from a single solar system when they are in the middle of the ISM is low.
Here we adopt the velocity of comet `Oumuamua \citep[$26.32 \pm 0.01$~km/s;][]{Mamajek2017} as typical of $\bar{v}_{\rm oc}$ for rocky objects crossing the ISM.
While the speed of interstellar objects may depend on their distance from the galactic center, we do not consider it here for simplicity. Combining $t_{\rm surv}$ and $\bar{v}_{\rm oc}$~yields a typical scale $\ell_{\rm surv}$ of a few parsecs, which matches well with the estimation of $\ell_{\rm surv} = 15-30 \pc$ in \citet{Grimaldi2021}. In the following section, we use a range of $\ell_{\rm surv} = 1-10^4 \pc$ to explore the parameter space and account for simple cases of \emph{directed} panspermia \citep[e.g., in case of $\ell_{\rm surv} \gtrsim 1 \kpc$; see also][]{Stapleton1930,Haldane1954,CrickOrgel1973}, while keeping in mind that $\ell_{\rm surv} \leq 30 \pc$ is likely more appropriate to the \emph{natural} case.

As the dependency of the damage weight $w_{\rm damage}$ to the travel distance $\ell \equiv |\vec{x}'_i - \vec{x}'_j|$ ($\vec{x}'_{i,j} \in \mathbb{V}_{i,j}$) is not precisely known, we consider three different models for $w_{\rm damage}$: (1) a \emph{sudden} damage model ({\tt sudden} hereafter), where spores stay alive at $\ell < \ell_{\rm surv}$ and suddenly die afterward; (2) a \emph{linear} damage model ({\tt lin} hereafter), in which the population linearly decreases over time until it becomes zero at $\ell = \ell_{\rm surv}$; and (3) an \emph{exponential} damage model ({\tt exp} hereafter), where the population of viable spores exponentially decreases over time, assuming that the survival rate over a fixed time period is constant:
\begin{equation}
w_{\rm damage}\left( \frac{\ell}{\ell_{\rm surv}} \right) = \left\{
\begin{array}{ll}
\mathcal{H}\left( 1 - {\ell}/{\ell_{\rm surv}} \right) & ({\tt sudden}) \\
\max \left\{ 1 - {\ell}/{\ell_{\rm surv}}, 0 \right\} & ({\tt lin}) \\
\exp \left( - {\ell}/{\ell_{\rm surv}} \right) & ({\tt exp}) \\
\end{array} \right.
\label{eq:damage_model}
\end{equation}
Finally, we also consider a fourth damage model, {\tt noEsc}, which is identical to {\tt sudden} where $f_{\rm esc}  = 1$ for all star particles, in order to understand the dependency of the panspermia probability on $f_{\rm esc}$. We also note that, unlike {\tt sudden} and {\tt lin}, 37\% of the spores are still alive at $\ell = \ell_{\rm surv}$ in {\tt exp}.
Therefore, one should be careful when comparing the dependency of the panspermia probability on $\ell_{\rm surv}$ between {\tt exp} and others.

\subsection{Numerical Formalism}\label{sec:pans_form}

Summarizing the above subsections, one can rewrite Eqs.~(\ref{eq:prob_onebyone}-\ref{eq:prob_onebyone2}) as follows:
\begin{equation}
f_{\rm pp}(\vec{x}_j) = \sum_i \mathcal{F}(\vec{x}_i, \vec{x}_j) \, ,
\label{eq:prob_new}
\end{equation}
where
\begin{multline}
\mathcal{F}(\vec{x}_i, \vec{x}_j) \propto f_{\rm cradle}(\vec{x}_i) f_{\rm esc}(\vec{x}_i) f_{\rm target}(\vec{x}_j) \\
\times \int_{\ell_{\rm min}}^{\ell_{\rm max}} \!\! \d\ell \, F(\ell|R_i,R_j,D_{ij}) w_{\rm damage}\left( \frac{\ell}{\ell_{\rm surv}} \right) \, . \label{eq:prob_frac}
\end{multline}
Since all timescales in Eqs. (\ref{eq:prob_onebyone}-\ref{eq:prob_onebyone2}) are smaller than the dynamic timescale of the Galaxy, we neglect time-dependent terms in the original equations. 
Here, $D_{ij} \equiv |\vec{x}_i - \vec{x}_j|$ is the distance between the centers of two star particles, considered as volumes homogeneously filled with stars. Also, $\ell_{\rm min} \equiv \max \left\{0, D_{ij} - R_i - R_j\right\}$ and $\ell_{\rm max} \equiv \min \left\{\alpha_{\rm trunc} \ell_{\rm surv}, D_{ij} + R_i + R_j\right\}$ are the minimum and maximum values of $\ell$ during the integration, respectively. 
A truncation rate $\alpha_{\rm trunc}$ is defined as the minimum value that satisfies $w_{\rm damage}(\alpha_{\rm trunc}) = 0$, whose value is 1 for {\tt sudden} and {\tt lin}.
For {\tt exp}, we manually set $\alpha_{\rm trunc}^{\tt exp} \gg 1$ for a fast calculation.

$F(\ell|R_i,R_j,D_{ij})$ is the probability of having $|\vec{x}'_i - \vec{x}'_j| = \ell$ given $R_i$, $R_j$, and $D_{ij}$, and assuming that the timescale for panspermia is much less than the evolution timescale of the galaxy:
\begin{multline}
F(\ell | R_i, R_j, D_{ij}) \\
= \frac{3}{R_i^3} \int \d D \, D^2 f_{\rm sp}(D|D_{ij},R_i) f_{\rm sp}(\ell |D,R_j) \, ,
\end{multline}
where
\begin{align}
f_{\rm sp}(d|D,R) &= \left\{ \begin{array}{ll} 0 & {\rm at\,} d > D+R \\
\mathcal{H}(R-D) & {\rm at\,} d < |D-R| \\
\begin{displaystyle}\frac{R^2 - (d-D)^2}{4Dd}\end{displaystyle} & {\rm otherwise} \end{array} \right.
\end{align}
is the fraction of a spherical surface with radius $d$ that belongs to the volume of another sphere of radius $R$ at distance $D$. 

From Eq.~(\ref{eq:prob_new}), one can also estimate a probability that seeds spreading from a given star particle are successfully transplanted in another (hereafter, the \emph{successful transplantation probability}):
\begin{equation}
f_{\rm stp}(\vec{x}_i) = \sum_j \mathcal{F}(\vec{x}_i, \vec{x}_j) \, .
\label{eq:prob_source}
\end{equation}

Since several factors in Eqs.~(\ref{eq:prob_onebyone}) \& (\ref{eq:prob_onebyone2}) remain unknown, Eqs.~(\ref{eq:prob_new}) \& (\ref{eq:prob_source}) thus only yield relative, rather than absolute, values of the panspermia probability.
However, since our main focus is a relative comparison between habitability and panspermia probability, we hereafter normalize $f_{\rm pp}$ and $f_{\rm stp}$ so that their sum over {\tt g15784} is same to the sum of the habitability, that is, 
\begin{equation}
\sum_{\vec{x}} f_{\rm pp}(\vec{x}) = \sum_{\vec{x}} f_{\rm stp}(\vec{x}) = \sum_{\vec{x}} f_{\rm hab}(\vec{x}) \, .
\label{eq:fpp_fstp_norm}
\end{equation}


\section{Results}\label{sec:res}

\subsection{Panspermia Probability \& Successful Transplantation Probability}\label{sec:res_fpp_fstp}

\begin{figure*}[bt]
\centering
\includegraphics[width=0.95\textwidth]{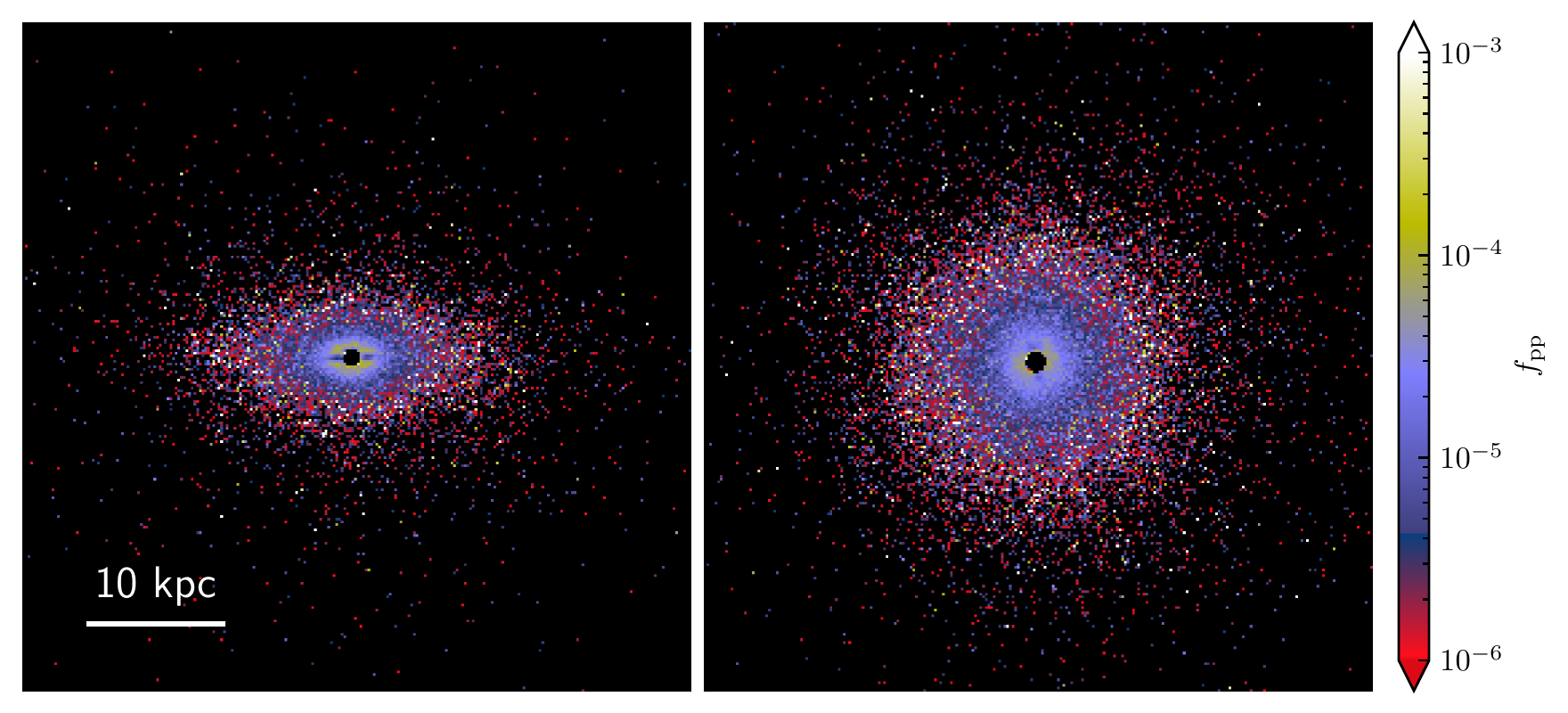}
\includegraphics[width=0.95\textwidth]{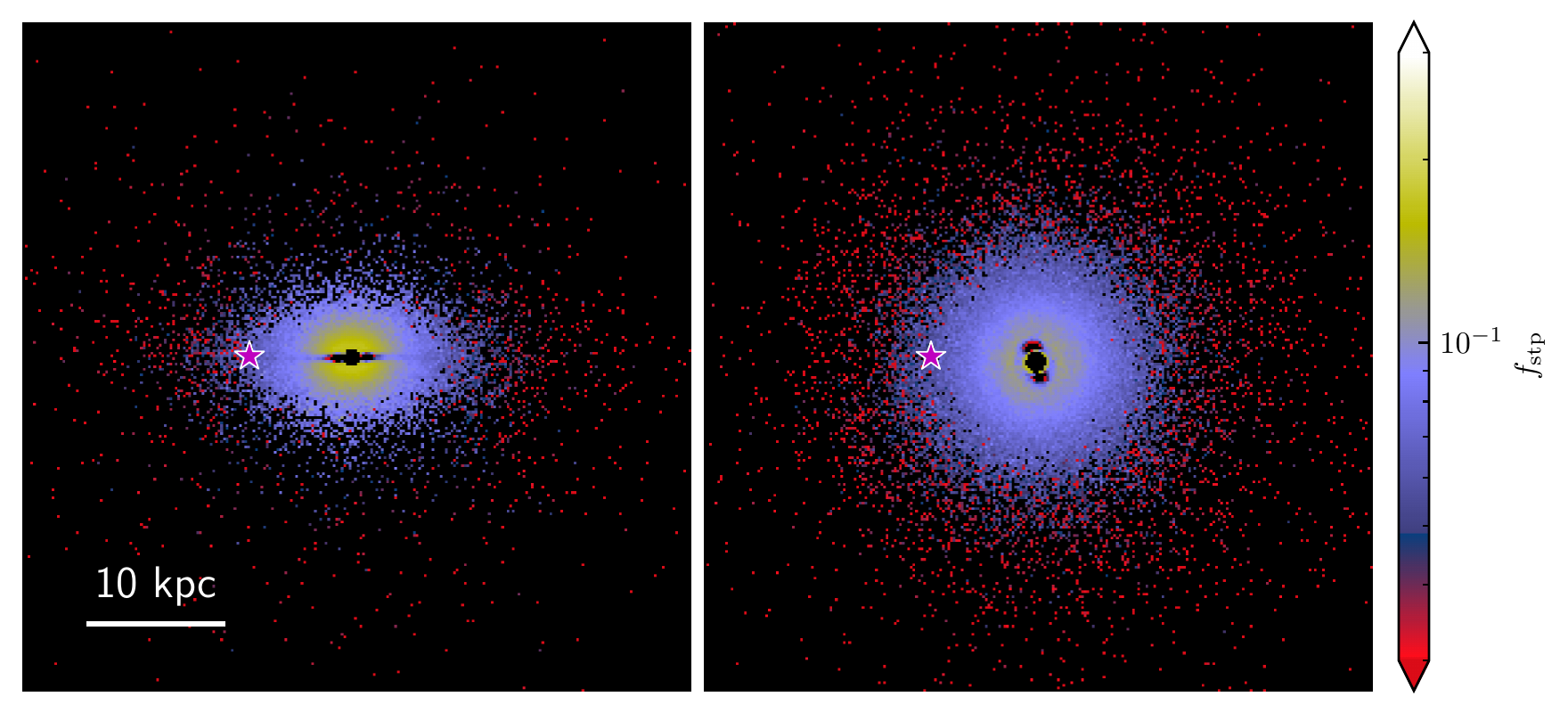}
\caption{Panspermia probability ($f_{\rm pp}$; \emph{top}) and successful transplantation probability ($f_{\rm stp}$; \emph{bottom}) at $z = 0$ in {\tt g15784}, assuming the exponential damage model ({\tt exp}) with $\ell_{\rm surv} = 100\pc$. The field of view, slice width, and the magenta star are set to be the same as Fig.~\ref{fig:fhab}.}
\label{fig:fpp_fstp_slice}
\end{figure*}

\begin{figure}[bt]
\centering
\includegraphics[width=0.495\textwidth]{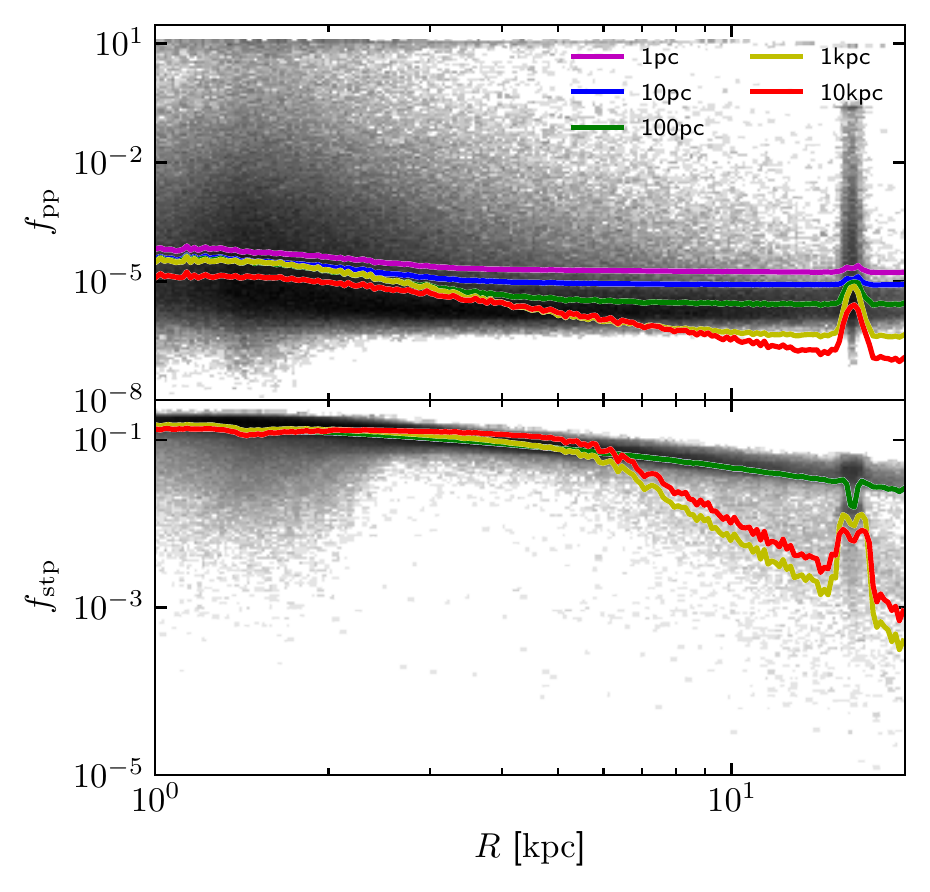}
\caption{Panspermia probability ($f_{\rm pp}$; \emph{top}) and successful transplantation probability ($f_{\rm stp}$; \emph{bottom}) at $z = 0$ in {\tt g15784} as a function of galactocentric radius. The joint distribution is drawn in logarithmic scale and grayscale, assuming {\tt exp} and $\ell_{\rm surv} = 100\pc$. On the other hand, the median probability for different values of $\ell_{\rm surv}$, for the same damage model, are shown by colored lines.}
\label{fig:fpp_fstp_radial}
\end{figure}

\begin{figure}[bt]
\centering
\includegraphics[width=0.495\textwidth]{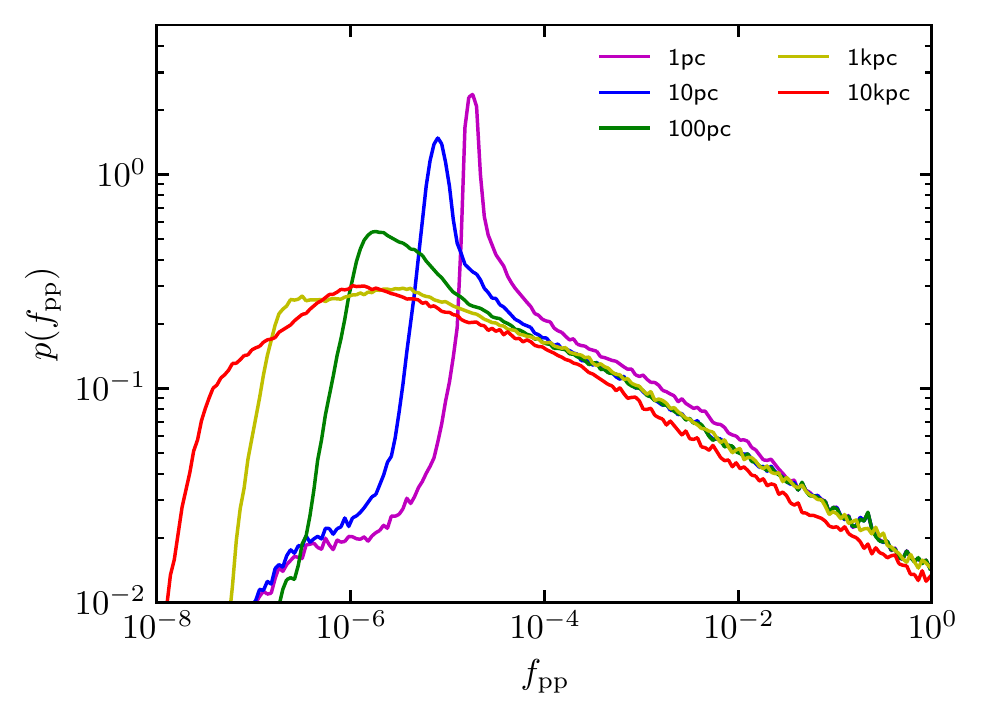}
\includegraphics[width=0.495\textwidth]{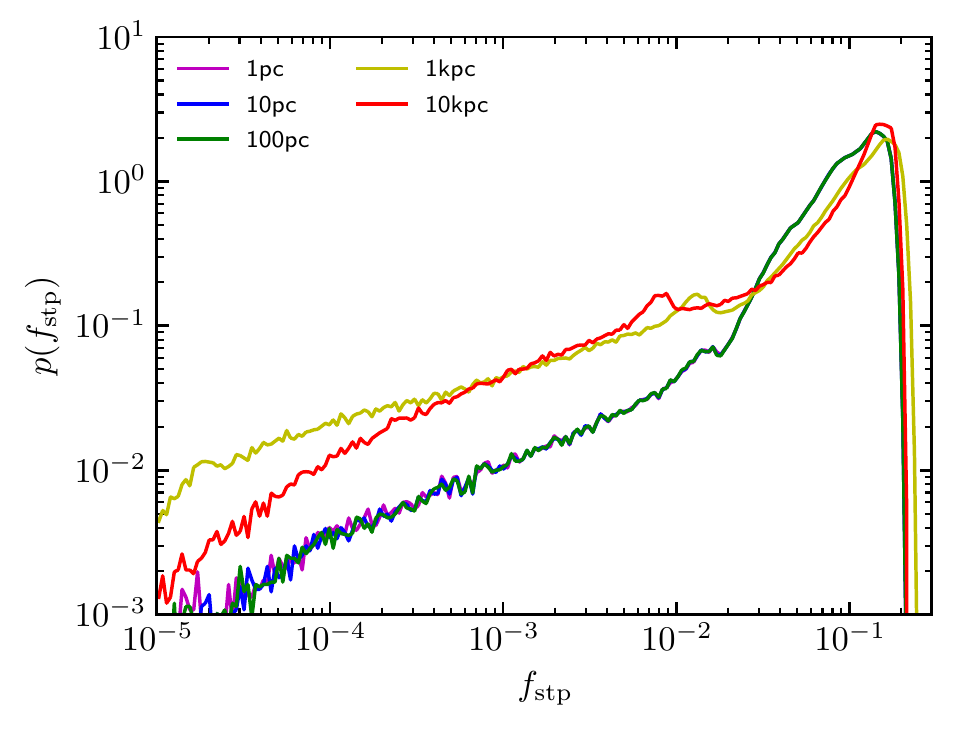}
\caption{Probability distributions of the panspermia ($f_{\rm pp}$; \emph{top}) and successful transplantation probabilities ($f_{\rm stp}$; \emph{bottom}) in {\tt g15784} at $z = 0$.
Both $f_{\rm pp}$ and $f_{\rm stp}$ are calculated by assuming exponential damage model ({\tt exp}) and various travel distance scales (\emph{colors}).}
\label{fig:fpp_fstp_1D}
\end{figure}

Figs.~\ref{fig:fpp_fstp_slice}--\ref{fig:fpp_fstp_1D} show the spatial and probability distributions of the panspermia probability and successful transplantation probability in {\tt g15784} at $z = 0$, normalized following Eq.~(\ref{eq:fpp_fstp_norm}).
Similarly to the three habitability levels, star particles in the {\tt CentralDisk} have very low $f_{\rm pp}$ and $f_{\rm stp}$.
In the {\tt DiskHalo} region, both $f_{\rm pp}$ and $f_{\rm stp}$ tend to have higher values at lower galactocentric radius ($R$) and tangential distance from the galactic plane ($b$).
However, while $f_{\rm pp}$ covers a wide range of 7 orders of magnitude and shows a somewhat mixed distribution  at $R > 5\kpc$, the successful transplantation probability $f_{\rm stp}$ shows a narrower range (3 orders of magnitude) and stronger dependency on both $R$ and $b$, in the $\ell_{\rm surv} = 100\pc$ case.
For larger values of $\ell_{\rm surv}$, however, $f_{\rm pp}$ tends to show a stronger negative slope on a $R$-direction (upper panel of Fig.~\ref{fig:fpp_fstp_radial}).
Such a negative slope at high-$\ell_{\rm surv}$ might happen because of the difference, between low-$R$ (high-$\Sigma_\star$) and at high-$R$ (low-$\Sigma_\star$), in the number of sources that a given star particle can receive.

The probability distribution of $p(f_{\rm pp})$ peaks around $10^{-4}$ times the typical value of the habitabilities.
As the travel distance increases, the peak of $p(f_{\rm pp})$ shifts towards lower values, decreases, and widens. This is because a large $\ell_{\rm surv}$ allows seeds to reach distant star particles, which naturally increases  the total amount of successfully transplanted seeds in a given galaxy. However, since we normalize $f_{\rm pp}$ so that its sum over the galaxy is constant, the importance of each successfully transplanted seed decreases. On the other hand, the peak position of the probability distribution of normalized successful transplantation probability ($p(f_{\rm stp})$) remains similar regardless of the value of $\ell_{\rm surv}$.
Instead, the probability distribution spreads toward lower $f_{\rm stp}$ as the travel distance scale increases, especially when $\ell_{\rm surv} \gtrsim 100\pc$.

\begin{figure*}[bt]
\centering
\includegraphics[width=0.49\textwidth]{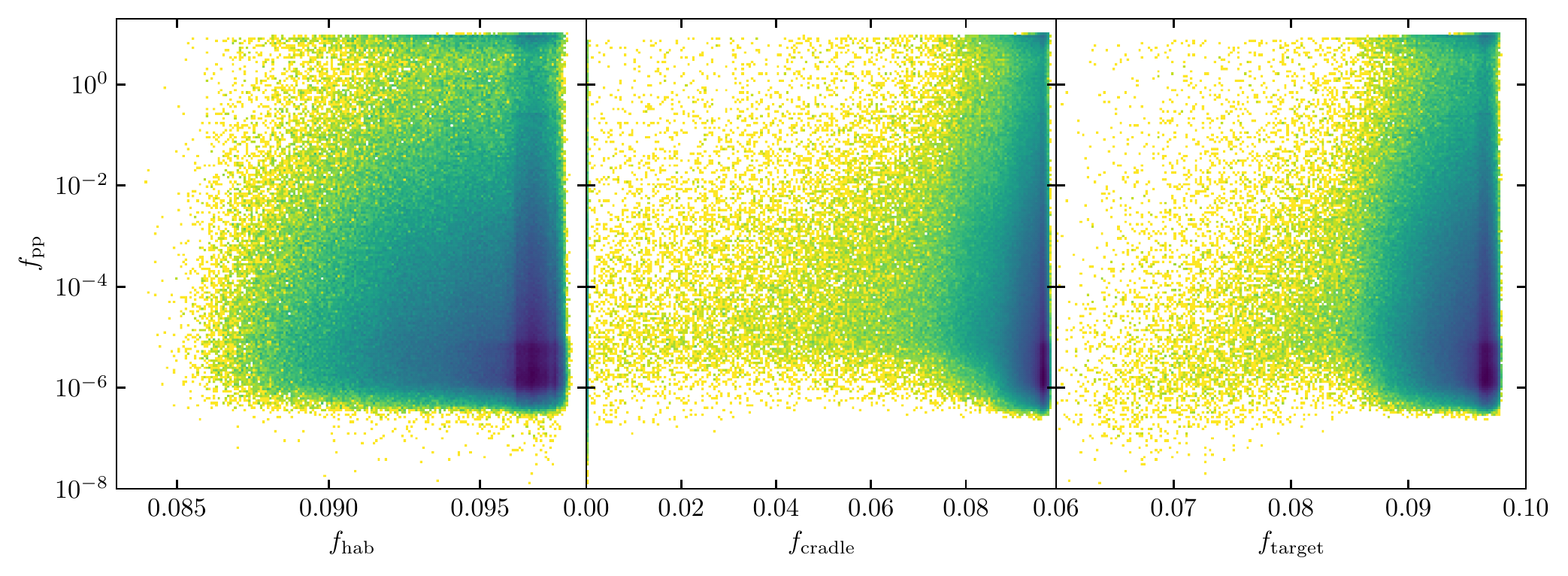}
\includegraphics[width=0.49\textwidth]{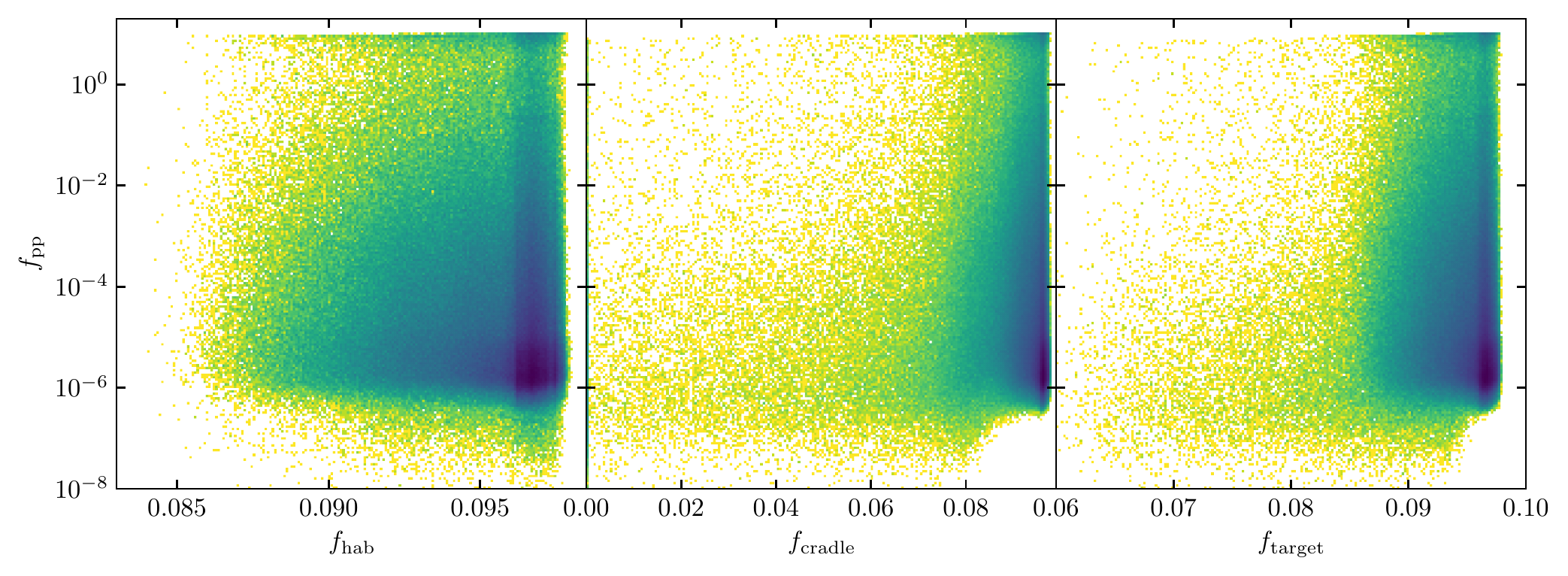}\\
\includegraphics[width=0.49\textwidth]{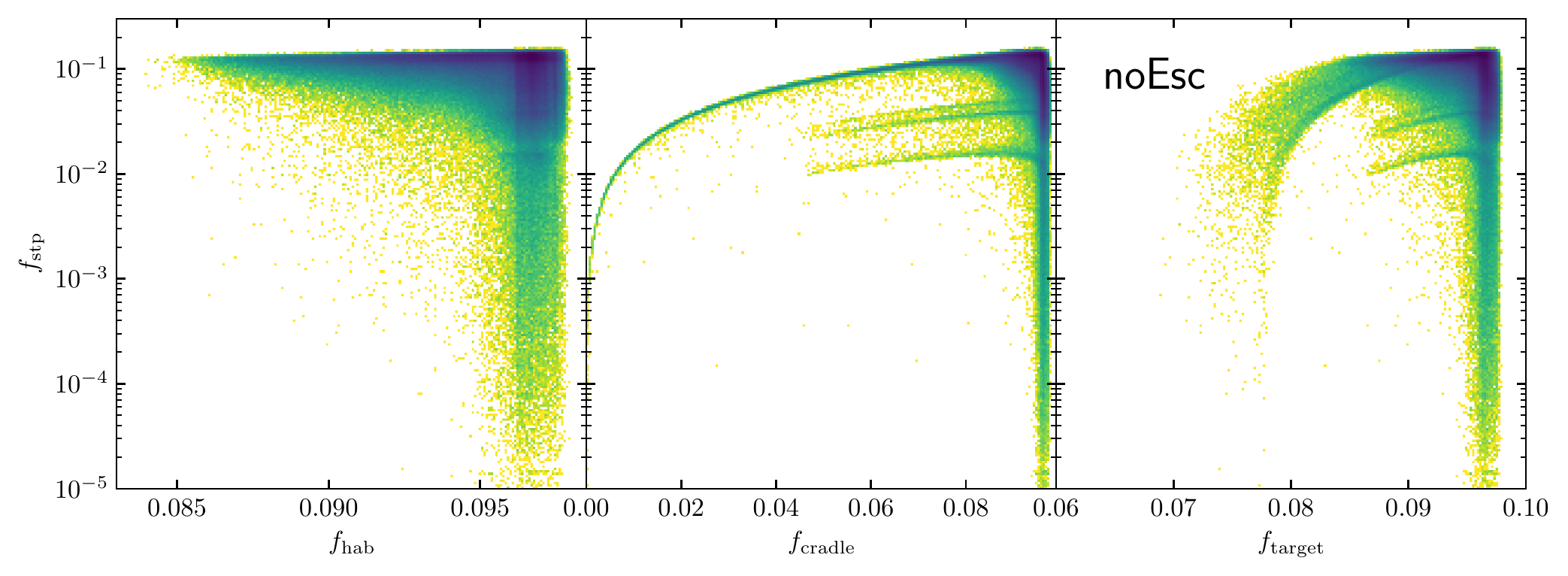}
\includegraphics[width=0.49\textwidth]{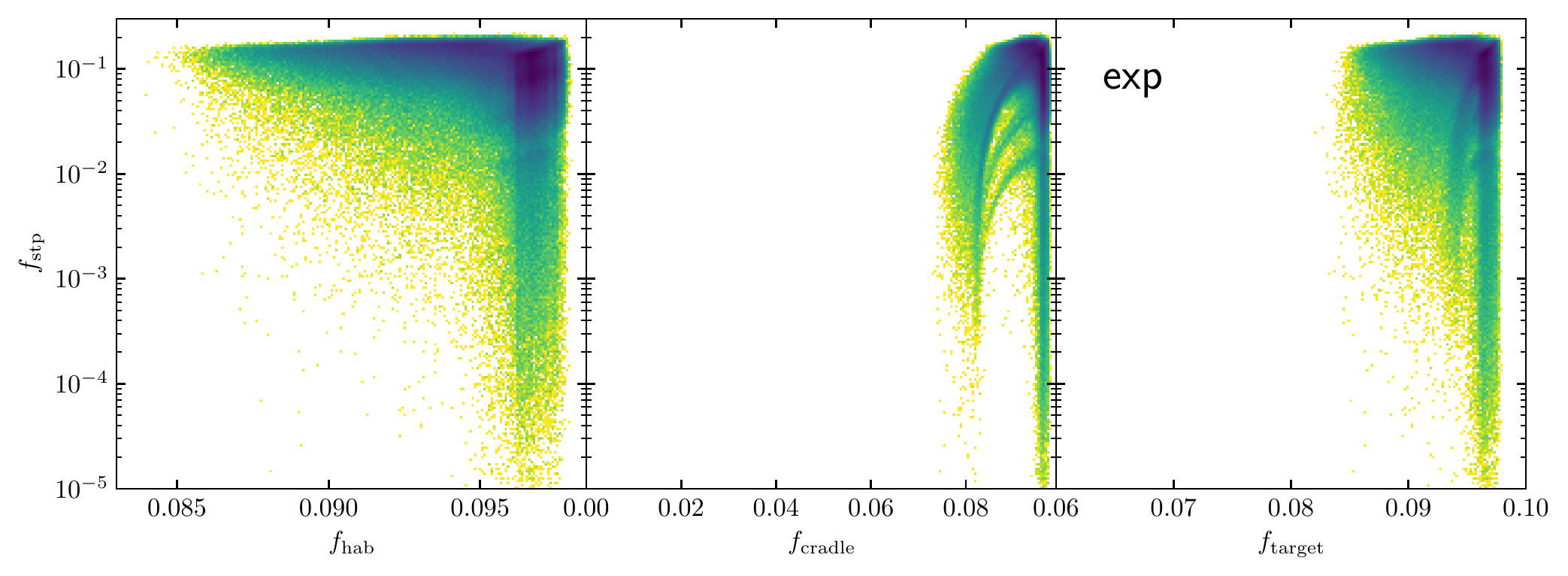}\\
\caption{The panspermia ($f_{\rm pp}$; \emph{top}) and successful transplantation ($f_{\rm stp}$; \emph{bottom}) probabilities as a function of three different types of habitabilities ($f_{\rm hab}$, $f_{\rm cradle}$, and $f_{\rm target}$) in {\tt g15784} at $z = 0$. 
\emph{Left}: Sudden damage model with $f_{\rm esc} = 1$ for all star particles ({\tt noEsc}) and $\ell_{\rm surv} = 100\pc$. 
\emph{Right}: Exponential damage model ({\tt exp}) with the same travel distance scale. Both the {\tt sudden} and {\tt lin} models give similar results to {\tt exp}.}
\label{fig:fpp_fstp_2D}
\end{figure*}

Fig.~\ref{fig:fpp_fstp_2D} shows $f_{\rm pp}$ and $f_{\rm stp}$ in {\tt g15784} at $z = 0$ as a function of the three habitability levels, for various damage models and $\ell_{\rm surv} = 100\pc$.
From Eqs.~(\ref{eq:prob_new}-\ref{eq:prob_frac}), the panspermia probability is 
\begin{multline}
f_{\rm pp}(\vec{x}_j) \propto f_{\rm target}(\vec{x}_j) \\
\times \sum_i f_{\rm cradle}(\vec{x}_i) f_{\rm esc}(\vec{x}_i) \mathbb{F}(R_i, R_j, D_{ij} |  \ell_{\rm surv}) \, , \label{eq:fpp_prop}
\end{multline}
where 
\begin{multline}
\mathbb{F}(R_i, R_j, D_{ij} | \ell_{\rm surv}) \\
\equiv \int_{\ell_{\rm min}}^{\ell_{\rm max}} \d \ell \, F(\ell | R_i, R_j, D_{ij}) w_{\rm damage}\left( \frac{\ell}{\ell_{\rm surv}} \right) \, , \label{eq:F_geometry}
\end{multline}
with $\mathbb{F}(\cdots)$ depending only on the geometry (volumes and relative position) of the source-target star particle pair. Consequently, while $\mathbb{F}$ may vary from particle to particle, $\sum_i \mathbb{F}$ is mostly determined by the local stellar density around the target star particle. Accordingly, while $\sum_i f_{\rm cradle}(\vec{x}_i) f_{\rm esc}(\vec{x}_i) \mathbb{F}(\cdots)$ is sensitive to both geometry and the baryonic properties of the target star particle, it can be approximated by a function of only $R_j$ in cases where $f_{\rm cradle} f_{\rm esc}$ does not vary much across all relevant source star particles. 
In other words, if the range of $f_{\rm target}$ in {\tt g15784} is significantly larger than that of $f_{\rm cradle} f_{\rm esc}$, then $f_{\rm pp} \propto f_{\rm target}$. On the other hand, if the range of $f_{\rm target}$ is similar or narrower than that of $f_{\rm cradle} f_{\rm esc}$, the above approximation cannot be used. As shown in Figs.~\ref{fig:habrad} \& \ref{fig:fhab_fesc}, $0.06 < f_{\rm target} < 0.1$ while $0 < f_{\rm cradle} f_{\rm esc} < 0.1$, even in the {\tt noEsc} with $f_{\rm esc} = 1$. Consequently, $f_{\rm pp}$ does not correlate clearly with habitability (upper panels of Fig.~\ref{fig:fpp_fstp_2D}). On the other hand, the successful transplantation probability can be written as
\begin{multline}
f_{\rm stp}(\vec{x}_i) \propto f_{\rm cradle}(\vec{x}_i) f_{\rm esc}(\vec{x}_i) \\
\times \sum_j f_{\rm target}(\vec{x}_j) \mathbb{F}(R_i, R_j, D_{ij} |  \ell_{\rm surv})
\end{multline}
and, in a similar way, we can assume that $\{ \mathbb{F}\}$ is mostly determined by $R_i$. Since the range of $f_{\rm cradle} f_{\rm esc}$ is significantly larger than that of $f_{\rm target}$, we can expect $f_{\rm stp} \propto f_{\rm cradle} f_{\rm esc}$. Indeed, Fig.~\ref{fig:fpp_fstp_2D} (bottom panels) shows strong correlations between $f_{\rm stp}$ and $f_{\rm cradle} f_{\rm esc}$, for different values $f_{\rm stp} / f_{\rm cradle} f_{\rm esc}$.
Most of the {\tt DiskHalo} population has high values of both $f_{\rm cradle}$ and $f_{\rm stp}$ ($(f_{\rm cradle}, f_{\rm stp}) \sim (0.09, 10^{-1})$), while values for {\tt CentralDisk}, on the other hand, are more widely spread, with a $f_{\rm stp} / f_{\rm esc} f_{\rm cradle} \simeq 2$ linear correlation (upper diagonal strip at the bottom panels of Fig.~\ref{fig:fpp_fstp_2D}). Finally, the lower diagonal strips in the bottom panel of Fig.~\ref{fig:fpp_fstp_2D} correspond to the {\tt Spheroids}.
Interestingly, there appears to be two distinct populations in the $f_{\rm cradle}$-$f_{\rm stp}$ plane, with $f_{\rm stp} / f_{\rm esc} f_{\rm cradle} \simeq 0.5$ and $0.2$, respectively. They disappear when $\ell_{\rm surv} \geq 10 \kpc$, suggesting that they originate from successful panspermia between {\tt Spheroids} and the main galaxy at low $\ell_{\rm surv}$.
The strength of correlation between $f_{\rm stp}$ and other two types of habitabilities depends on the correlation between those habitabilities and $f_{\rm cradle}$---as $f_{\rm target}$($f_{\rm hab}$) shows a mild(no apparent) correlation with $f_{\rm cradle}$, so does it with $f_{\rm stp}$ (see Fig.~\ref{fig:habrad}, for example).

At a fixed value of $\ell$, we find that neither the probability and joint distributions of $f_{\rm pp}$ and $f_{\rm stp}$ strongly depend on the damage model ({\tt sudden}, {\tt lin}, or {\tt exp}).
While the three damage models have different $w_{\rm damage}(\ell / \ell_{\rm surv})$, and therefore, different $\mathcal{F}(\vec{x}_i, \vec{x}_j)$, the normalization of $f_{\rm pp/stp}$ described above weakens the difference between them. That numerous star particles can affect both $f_{\rm pp}$ and $f_{\rm stp}$ further lessens it. On the other hand, there is significant difference between those three damage models and {\tt noEsc}, mostly due to differences in escape weight $f_{\rm esc}$.

\subsection{Panspermia Contributions}\label{sec:res_bestFrac}

By definition, both the $f_{\rm pp}$ and $f_{\rm stp}$ of a given star particle are affected by numerous source and target particles. This means that star particles with the same values of $f_{\rm pp}$ or $f_{\rm stp}$ may however have a completely different panspermia \emph{history}. For example, some star particles might have received similar amounts of seeds from numerous particles, while others might have received most of \emph{their seeds from only one} or a few dominant sources. As an attempt to quantify the panspermia history of individual particles,  we define the \emph{greatest panspermia contribution} as the highest single $f_{\rm pp}$ or $f_{\rm stp}$ contribution to the sum:\footnote{Ideally, all $\mathcal{F}(\vec{x}_i, \vec{x}_j)$ for the entire $N_\star^2$-pairs of possible routes should be considered. However, the large number of $N_\star = 746,659$ makes it impractical at the moment.
We consider the full list of panspermia contributions and their network in a future work.}
\begin{align}
\mathcal{F}_{\rm best,pp}(\vec{x}) &\equiv \max_{\vec{x}'} \mathcal{F}(\vec{x}',\vec{x}) \\
\mathcal{F}_{\rm best,stp}(\vec{x}) &\equiv \max_{\vec{x}'} \mathcal{F}(\vec{x},\vec{x}')
\end{align}

\begin{figure}[bt]
\centering
\includegraphics[width=0.495\textwidth]{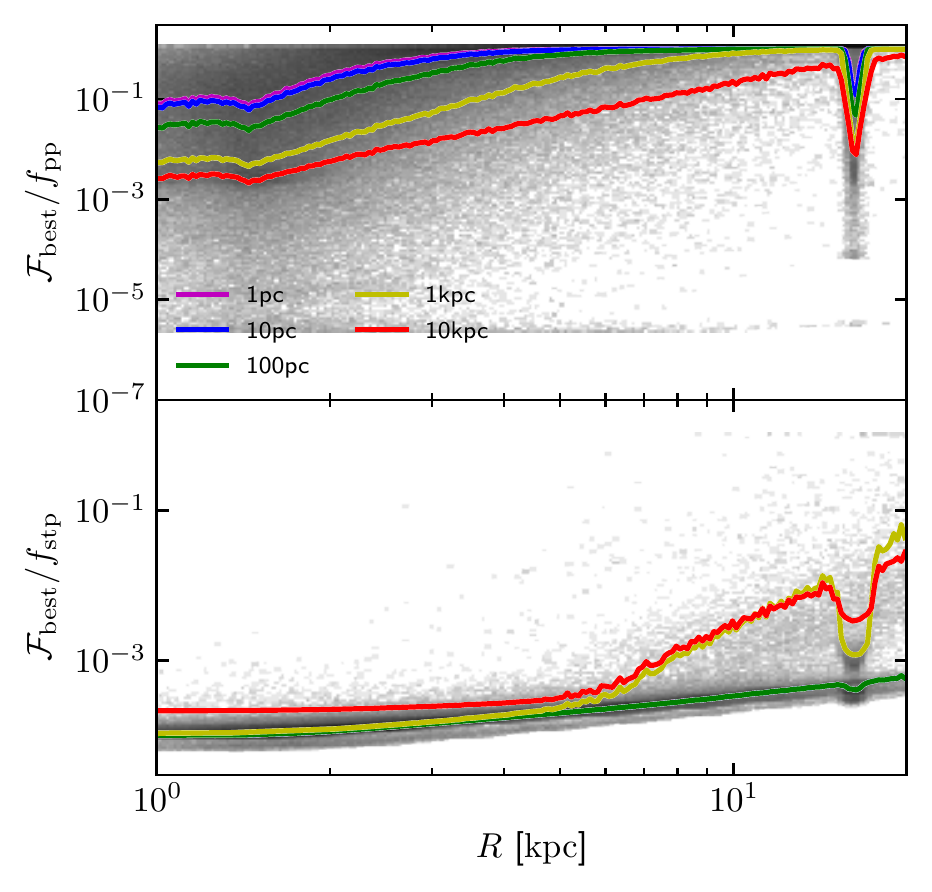}
\caption{Greatest contribution relative to the panspermia probability, $\mathcal{F}_{\rm best} / f_{\rm pp/stp}$), as a function of galactocentric radius. The colored lines show the median value, as in Fig.~\ref{fig:fpp_fstp_radial}.}
\label{fig:bestFrac_radial}
\end{figure}

Both $\mathcal{F}_{\rm best}$ occupy a narrow range of values and are nearly constant with galactocentric distance. On the other hand, as shown in Fig.~\ref{fig:bestFrac_radial}, their ratios to $f_{\rm pp}$ and $f_{\rm stp}$ show a clear radial trend: $\mathcal{F}_{\rm best} / f_{\rm pp}$ tends to be high in the {\tt DiskHalo}, increasing with radius up to values of $\sim$1, and low in both the {\tt CentralDisk} and {\tt Spheroids}. This implies that, while in the outer galactic disk and the halo the panspermia probability of a star particle is on average dominated by a single other particle (i.e., a one-to-one transmission), closer to the bulge and in the satellites the $f_{\rm pp}$ of star particles is a sum of contributions from multiple sources. In this case, the high local density more than offsets the lower habitability of particles in these regions. Likewise, $\mathcal{F}_{\rm best} / f_{\rm pp}$ decreases monotonically with increasing $\ell_{\rm surv}$, as the larger travel distance allows more source particles to contribute.
On the other hand, $\mathcal{F}_{\rm best} / f_{\rm stp}$ is somewhat insensitive to the travel distance until $\ell_{\rm surv} \gtrsim 1\kpc$, when other routes to target particles in the {\tt DiskHalo} and {\tt Spheroids} become possible.

\subsection{Self Panspermia in {\tt g15784}}\label{sec:res_self}

So far we considered all star particles in our calculation of $f_{\rm pp}$ and $f_{\rm stp}$, including cases where source and target particles are the same. This possibility of ``self panspermia'' accounts for the limited spatial and mass resolutions of MUGS, where each star particle ($6.3 \times 10^6 \Msun$) contains millions of stars. In this Section, we investigate the contribution of self panspermia to the panspermia probability. 

\begin{figure}[bt]
\centering
\includegraphics[width=0.495\textwidth]{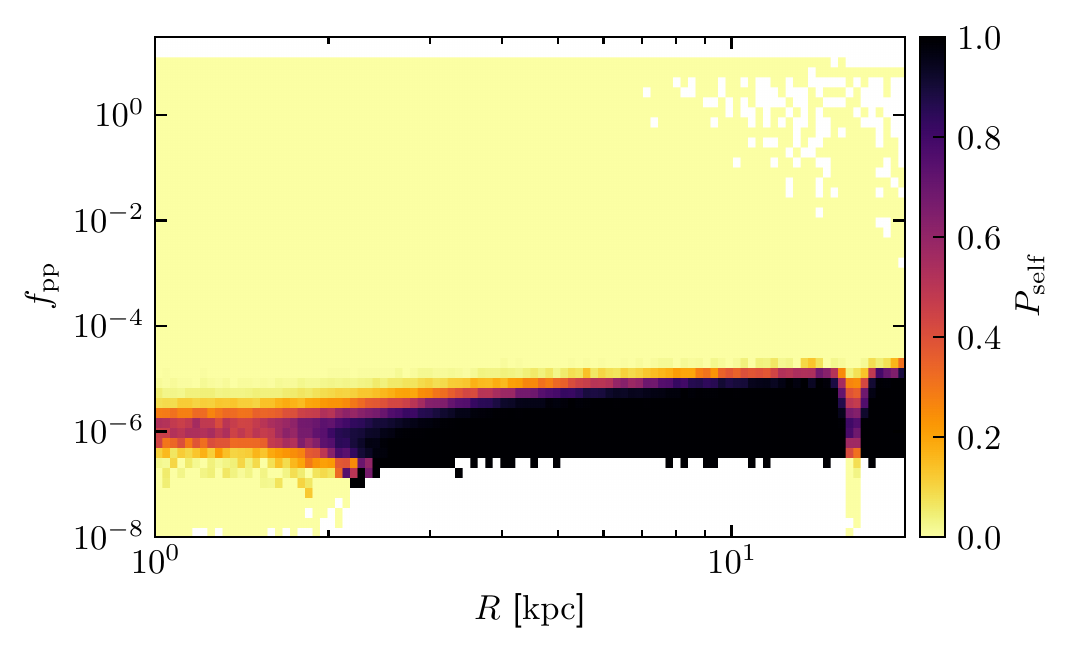}
\caption{The probability of \emph{self panspermia} ($P_{\rm self}$), where $\mathcal{F}(\vec{x},\vec{x}) > f_{\rm pp/stp}(\vec{x})/2$, as a function of $R$ and $f_{\rm pp}$}
\label{fig:self_radial}
\end{figure}

We consider that a given star particle undergoes self panspermia if the panspermia contribution from/to the same star particle is greater than the sum of the contributions from/to the all other star particles, i.e., star particle $i$ undergoes self panspermia when $\mathcal{F}(\vec{x}_i,\vec{x}_i) > f_{\rm pp/stp} (\vec{x}_i) / 2$.
Fig.~\ref{fig:self_radial} shows the distribution of self panspermia in {\tt g15784} at $z = 0$ as a function of galactocentric radius and $f_{\rm pp}$, assuming {\tt exp} with $\ell_{\rm surv} = 100\pc$. Self panspermia starts to dominate at $R > 2 \kpc$, where most of the {\tt DiskHalo} population exists, and for particles with panspermia probabilities at or below the median value (here $f_{\rm pp}\sim10^{-5}$). This threshold decreases by two orders of magnitude when $\ell_{\rm surv}$ increases from $1\pc$ to $10\kpc$. Conversely, no self panspermia occurs in star particles with values of $f_{\rm pp}$ $\gtrsim 10^2$ times greater than the median.
On the other hand, star particles at lower galactocentric radii, both in {\tt CentralDisk} and {\tt DiskHalo}, tend to undergo less self panspermia because the high stellar densities near the galactic center may prevent the panspermia process from being governed by a single route. Finally, we find almost no cases of self panspermia for the successful transplantation probability ($f_{\rm stp}$), except in except in the outer, underdense region of the {\tt DiskHalo}.

\begin{figure}
\centering

\includegraphics[width=0.495\textwidth]{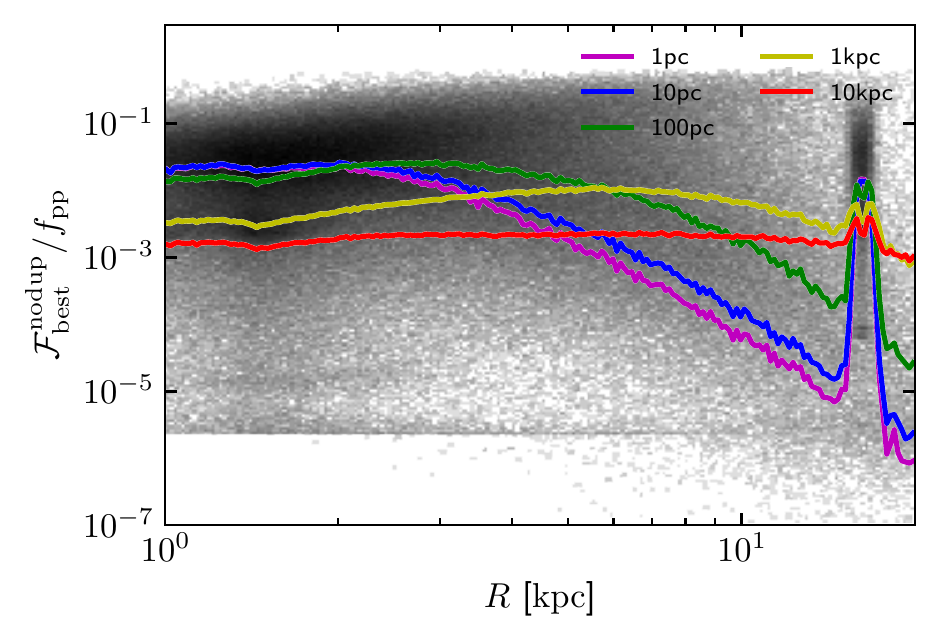}
\caption{Same as Fig.~\ref{fig:bestFrac_radial} (\emph{top}), but for $\mathcal{F}_{\rm best}^{\rm nodup} / f_{\rm pp}$, i.e., without the contribution of self panspermia.}
\label{fig:bestFrac_nodup}
\end{figure}

Having estimated self panspermia, we subtract it from $\mathcal{F}_{\rm best}$:
\begin{align}
\mathcal{F}_{\rm best,pp}^{\rm nodup}(\vec{x}) &\equiv \max_{\vec{x}' \neq \vec{x}} \mathcal{F}(\vec{x}',\vec{x}) \\
\mathcal{F}_{\rm best,stp}^{\rm nodup}(\vec{x}) &\equiv \max_{\vec{x}' \neq \vec{x}} \mathcal{F}(\vec{x},\vec{x}') \, ,
\end{align}
with Fig.~\ref{fig:bestFrac_nodup} showing the behavior of $\mathcal{F}_{\rm best}^{\rm nodup} / f_{\rm pp/stp}$.
Contrary to the full case (Fig.~\ref{fig:bestFrac_radial}), $\mathcal{F}_{\rm best}^{\rm nodup} / f_{\rm pp}$ is strongly suppressed at $R > 2\kpc$, in agreement with Fig.~\ref{fig:self_radial}. This implies that, in the  {\tt DiskHalo}, most of $f_{\rm pp}$ is self panspermia, but also that the number of sources contributing to non-self panspermia increases with galactocentric radius. On the other hand, the latter (i.e.,panspermia from other source particles) dominates the {\tt CentralDisk} and {\tt Spheroids}.
We also note that, while the fraction of star particles dominated by self panspermia is large, these tend to have lower values of $f_{\rm pp}$, as discussed above and therefore contribute less to galactic panspermia.

\subsection{Panspermia vs. Prebiotic Evolution}

In this work we have eschewed a key question that implicitly motivates it, namely: \emph{which could be the dominant source of life on habitable planets in the galaxy, in-situ evolution or panspermia ?}
Within the methodology used in this paper, assuming that $f_{\rm oc}$ does not depend on, e.g., metallicity and that the timescales of prebiotic evolution and transmission are short compared to the age difference between particles, this could be parameterized proportionally to the ratio between the sums of panspermia probability and of natural habitability. Under these conditions, then, panspermia would take precedence when the habitability of a given particle is lower than the sum of habitabilities of neighboring particles weighted by escape fraction and distance.

In practice, however, since $f_{\rm pp/stp}$ have to be arbitrarily scaled due to the presence of undetermined parameters, this question cannot be answered quantitatively. The actual, \emph{unnormalized} panspermia probability can be defined as $\hat{f}_{\rm pp,stp} = \mathcal{R} \times f_{\rm pp,stp}$, with the true value of $\mathcal{R}$ being unknown. Varying $\mathcal{R}$ for the probability of $\hat{f}_{\rm pp} > f_{\rm hab}$ and $\hat{f}_{\rm stp} > f_{\rm hab}$, we find that matching the na\"ive expectation that $P(\mathcal{R}) \simeq \mathcal{R} / (\mathcal{R} + 1)$ would require $\mathcal{R} \gtrsim 10^{3.5}$ and $\mathcal{R} \gtrsim 1$, respectively (for more detail, see Appendix~\ref{appendix:fhat}). Fig.~\ref{fig:high_pp_radial} shows the radial distribution of star particles with $\hat{f}_{\rm pp/stp} > f_{\rm hab}$ for various values of $\mathcal{R}$ which, for $\mathcal{R} > 10^{-2}$, follows a similar slope to $R$-$\Sigma_\star$ in Fig.~\ref{fig:g15784_profile}. This confirms that dense regions, such as the {\tt CentralDisk} and inner part of the {\tt DiskHalo}, have a higher chance of being seeded through panspermia.

\begin{figure}
\centering
\includegraphics[width=0.495\textwidth]{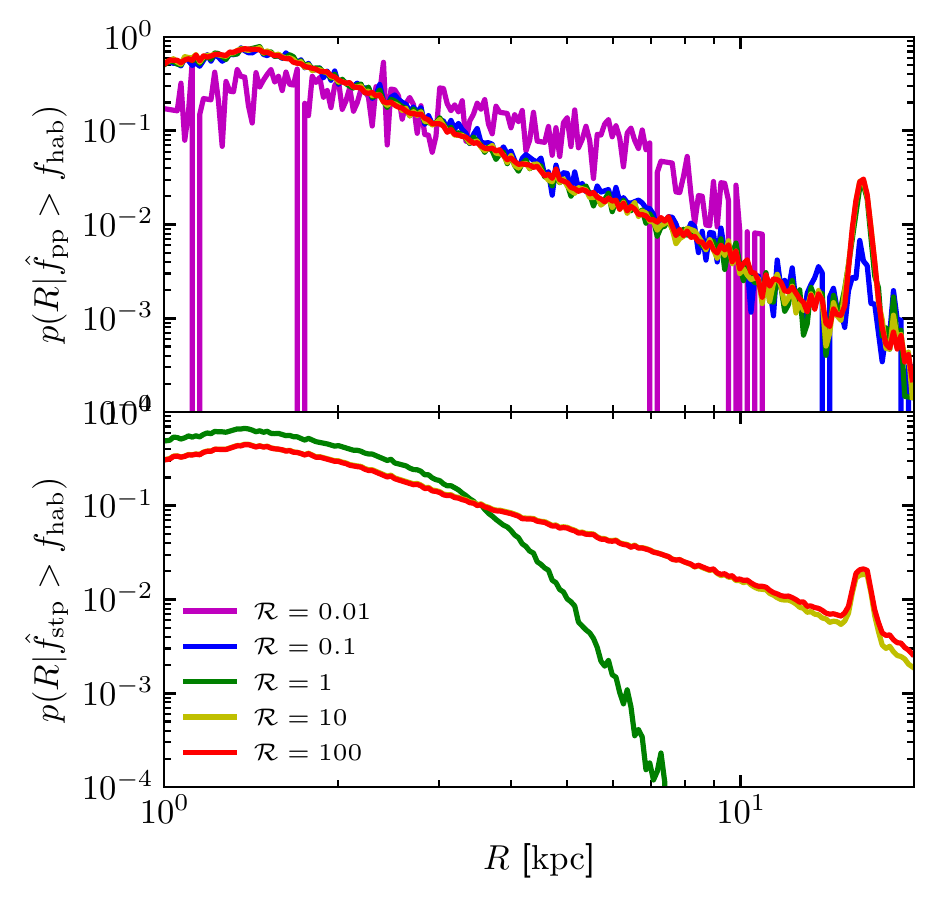}
\caption{Radial distribution of star particles with unnormalized panspermia  probabilities greater than the natural habitability ($p(R | \hat{f}_{\rm pp/stp} > f_{\rm hab})$) in {\tt g15784} at $z = 0$, for different values of $\mathcal{R}$. $p(R | \hat{f}_{\rm pp/stp} > f_{\rm hab})$ is normalized such that its integral over $R$ is one.
}
\label{fig:high_pp_radial}
\end{figure}


\section{Conclusions}\label{sec:concl}

In this paper, we have modeled the probability of panspermia (i.e., of successful material transfer between star systems) and its distribution in Milky Way-like galaxies, using a simulated object from the \emph{McMaster Unbiased Galaxy Simulations} (MUGS). To compute panspermia probabilities, we have expanded on the formalism presented in \citetalias{Gobat2016}, adding models for the ejection of spores from planets, their escape, transit, and in-transit damage. Our conclusions are summarized as follows:
\begin{enumerate}
\item While the median habitability of increases with galactocentric radius, the probability for panspermia behaves inversely, being likelier in the central regions of the bulge ($R = 1-4 \kpc$), as in the compact dwarf spheroids orbiting the simulated main galaxy. This is mostly due to higher stellar densities, which counterbalances their lower habitability. On the other hand, the panspermia probability is low in the central \emph{disk}, owing to a lower escape fraction due to metallicity and higher supernova rates. 
In dense regions, many source particles can contribute to panspermia, whereas in the outer disk and halo the panspermia probability is typically dominated by one or, at most, a few source star particles. Unlike natural habitability, whose value varies by only $\sim 5\%$ throughout the galaxy, the panspermia probability has a wide dynamic range of several orders of magnitudes.
\item There is no clear correlation between the panspermia probability and the habitability of the receiving particle, mainly because the former, especially at high values, is affected by numerous source star particles. On the other hand, it is strongly correlated with the habitability of the source particle, with several distinct stellar populations standing out, corresponding to those in the central disk, outside of the central disk, and in the satellites.
\item Finally, although this cannot be precisely quantified at the moment, we expect the process of panspermia to be significantly less efficient at seeding planets than in-situ prebiotic evolution. For example, even in a saturated case where the total panspermia probability is of the order of the total habitability in the galaxy, it would only dominate in 3\% of all star particles.
\end{enumerate}

Several caveats remain in our model: first, it includes several factors that we have regarded as unknown constants (e.g., the capture fraction of spores by target planets, the relation between habitability and the presence of life, the typical speed of interstellar objects, and the absolute value of escape fraction of the interstellar organic compounds from source planets). Our results are therefore naturally more qualitative than quantitative. Second, our calculation was done on a single simulation snapshot. That is, it assumes the spatial distribution to be static, while the actual Milky Way rotates and evolves. As such, these results only apply if the typical timescale for panspermia is much shorter than the dynamical timescale of a galaxy. Third, although we have used one of the best mock Milky-Way proxies available, some differences exist between the actual Milky Way and {\tt g15784}, which may lead to differences in panspermia probability. For example, our mock galaxy has a larger value of bulge-to-disc light ratio than the actual Milky Way \citep{Brook2012}, and the galactic bulge has been suggested to be well-suited for panspermia \citep[e.g.,][]{Chen2018,Balbi2020}. Finally, higher-resolution galaxy simulations with proper implementation of large-scale effects may provide a more realistic estimation of panspermia probability by, among other things, not having to account for self panspermia \citep[e.g.,][]{Schaye2015,Crain2015,Lee2021}.

\acknowledgements
The authors thank Michael Gowanlock and Heeseung Zoe for helpful discussions. We acknowledge Jeremy Bailin, Greg Stinson, Hugh Couchman and James Wadsley for carrying out the MUGS simulation and allowing us access to the data.
SEH was partly supported by Basic Science Research Program through the National Research Foundation of Korea funded by the Ministry of Education (2018\-R1\-A6\-A1\-A06\-024\-977).
SEH was also partly supported by the project \begin{CJK}{UTF8}{mj}우주거대구조를 이용한 암흑우주 연구\end{CJK} (``Understanding Dark Universe Using Large Scale Structure of the Universe''), funded by the Ministry of Science. 
ONS thanks the DIM ACAV+ for postdoctoral funding. 
The authors acknowledge the Korea Institute for Advanced Study for providing computing resources (KIAS Center for Advanced Computation Linux Cluster System).
Computational data were transferred through a high-speed network provided by the Korea Research Environment Open NETwork (KREONET).

\software{IDL, Matplotlib \citep{matplotlib2007}, NumPy/SciPy \citep{scipy2020}, Pandas \citep{pandas2010}}



\appendix
\restartappendixnumbering

\section{Escape fractions and weights}\label{appendix:wesc}

\begin{figure}
\centering
\includegraphics[width=0.495\textwidth]{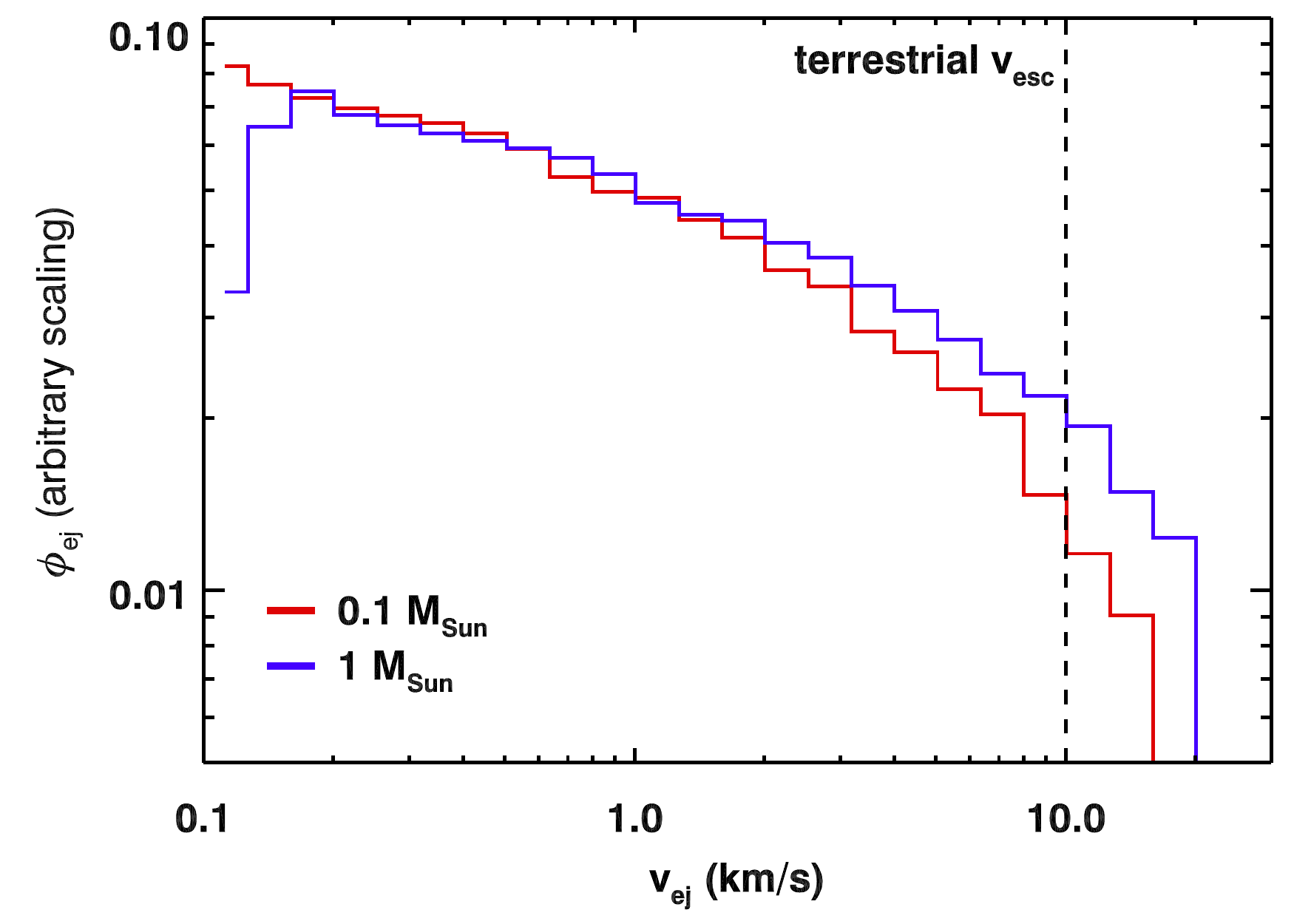}
\includegraphics[width=0.495\textwidth]{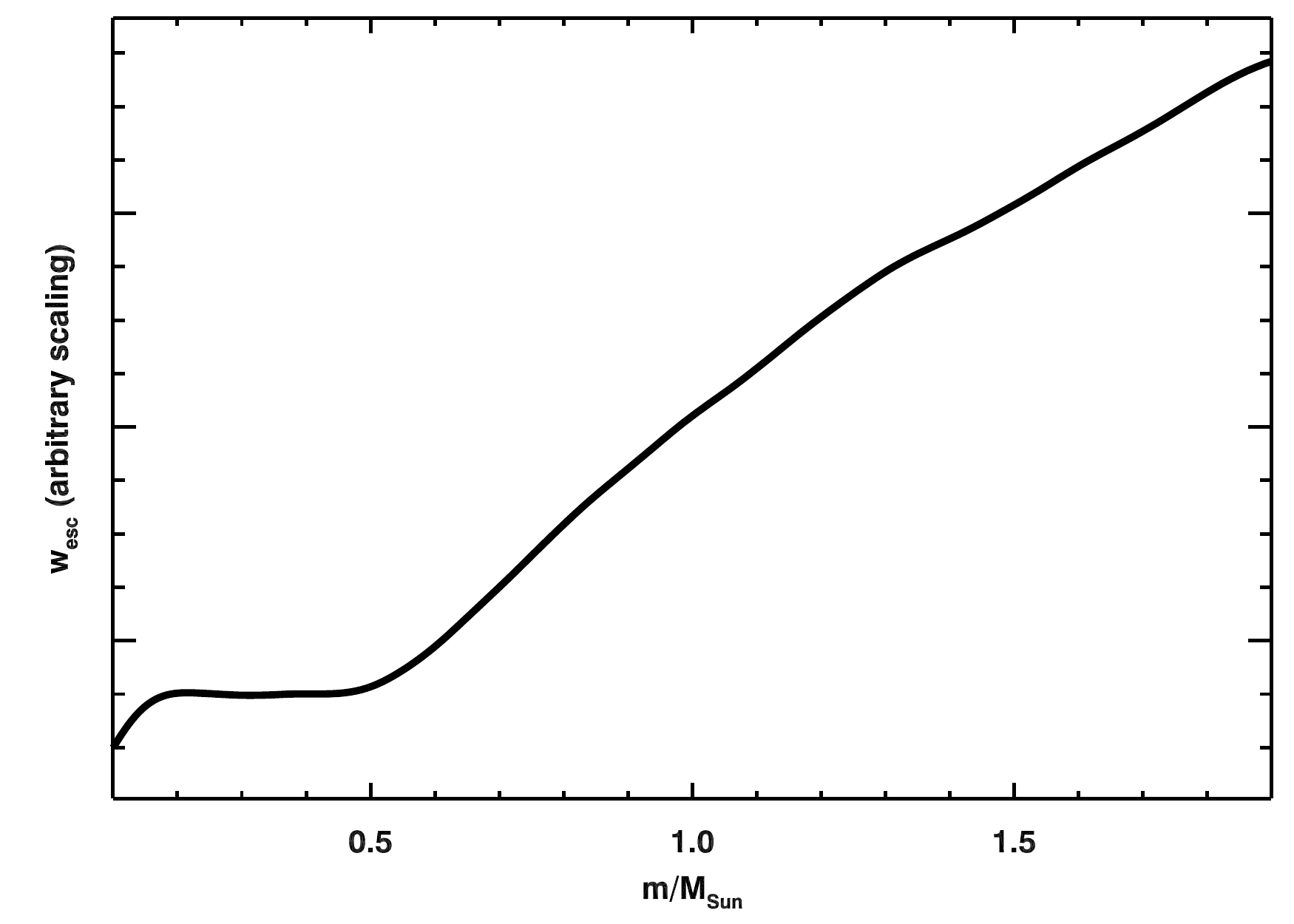}

\caption{{\it Top:} Distribution of ejecta velocities from an impactor population with 
$f(m)\propto m^{-1.6}$, for a planet within the HZ of a $0.1\Msun$ (red) and 
$1\Msun$ (blue) star, respectively. The dashed vertical line shows the approximate escape velocity of an Earth-mass planet, here set at $10 {\rm km/s}$ for convenience.
{\it Bottom:} relative fraction of escaping material $w_{\rm esc}$ as a function of host star mass, assuming that the mass of HZ terrestrial planets does not vary with host star mass. Only the mass range for which the HZ is defined \citep{Kopparapu2013} is shown.
}
\label{fig:ejecta1}
\end{figure}

\begin{figure}
\centering
\includegraphics[width=0.45\textwidth]{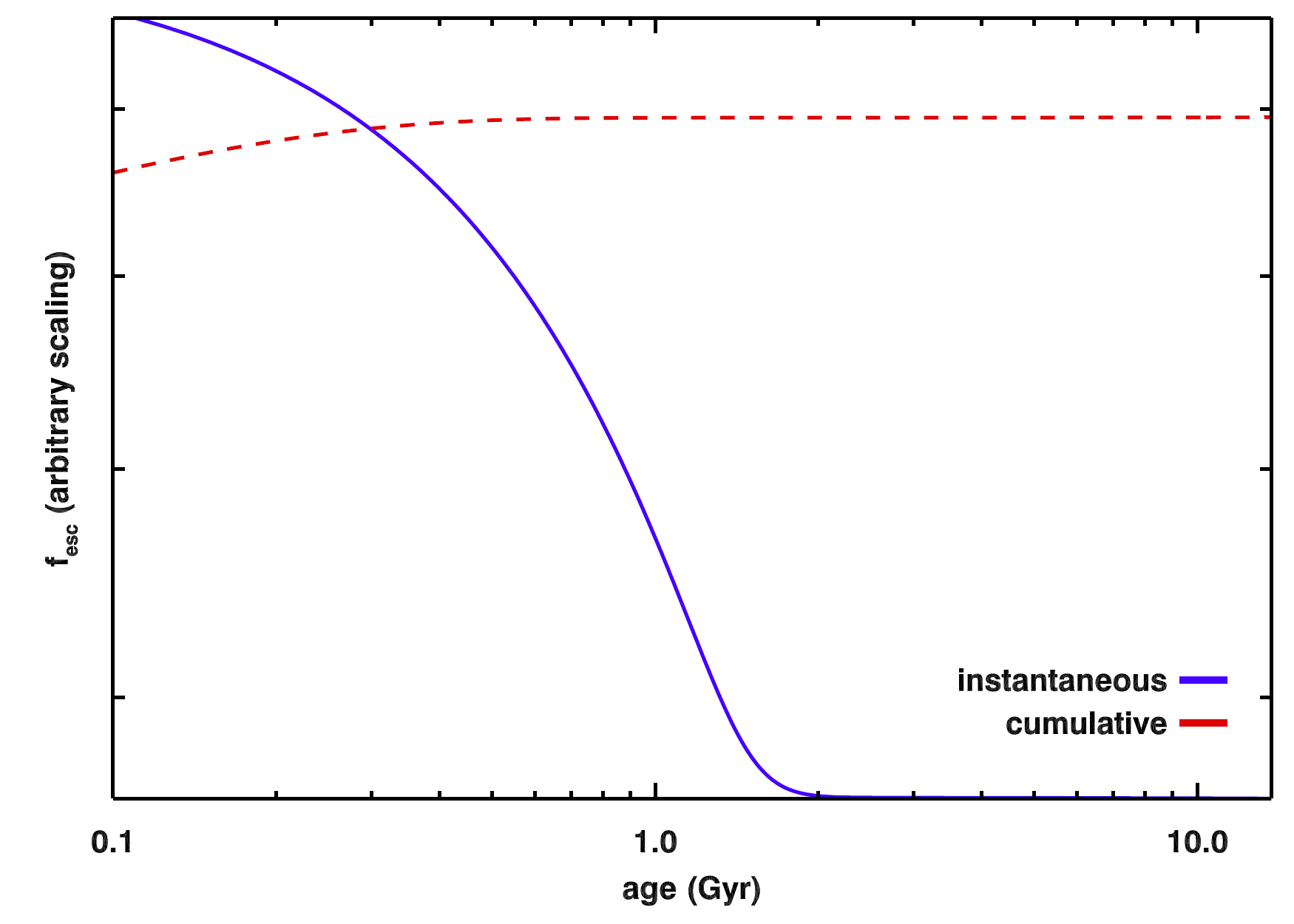}
\caption{Escape fractions for star particles with a \citetalias{Kroupa1993} IMF, as a function of their age. The solid blue curve shows the instantaneous $f_{\rm esc}$ and the dashed red one the cumulative $f_{\rm esc}$ over the lifetime of the star particle.
}
\label{fig:ejecta2}
\end{figure}

\begin{figure}
\centering
\includegraphics[width=0.48\textwidth]{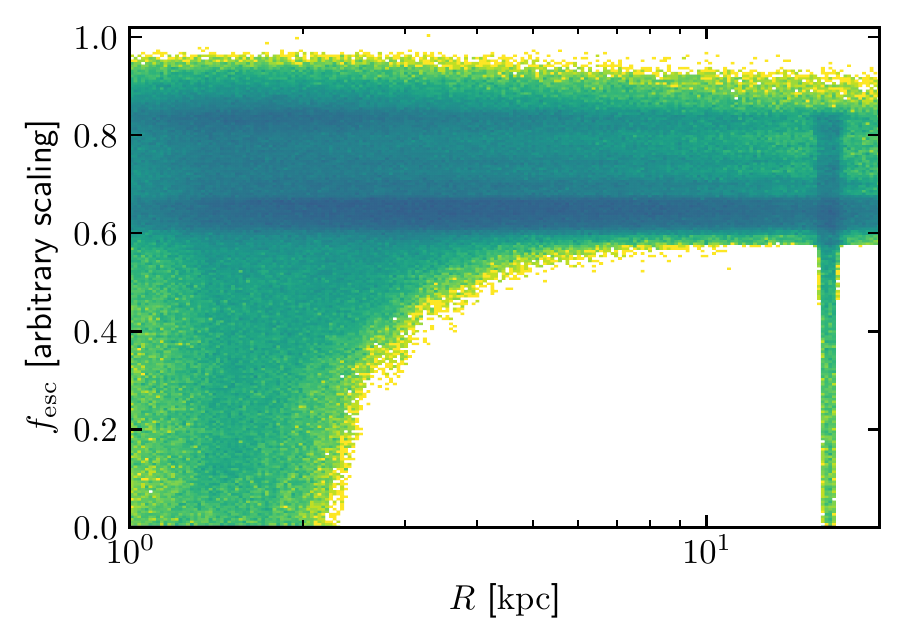}
\caption{The escape weight ($f_{\rm esc}$) as a function of galactocentric radius in {\tt g15784} at $z = 0$, drawn in a logarithmic scale.
Note that the escape weight is shown in artibrary unit.}
\label{fig:fesc_radial}
\end{figure}

\begin{figure*}[t]
\plotone{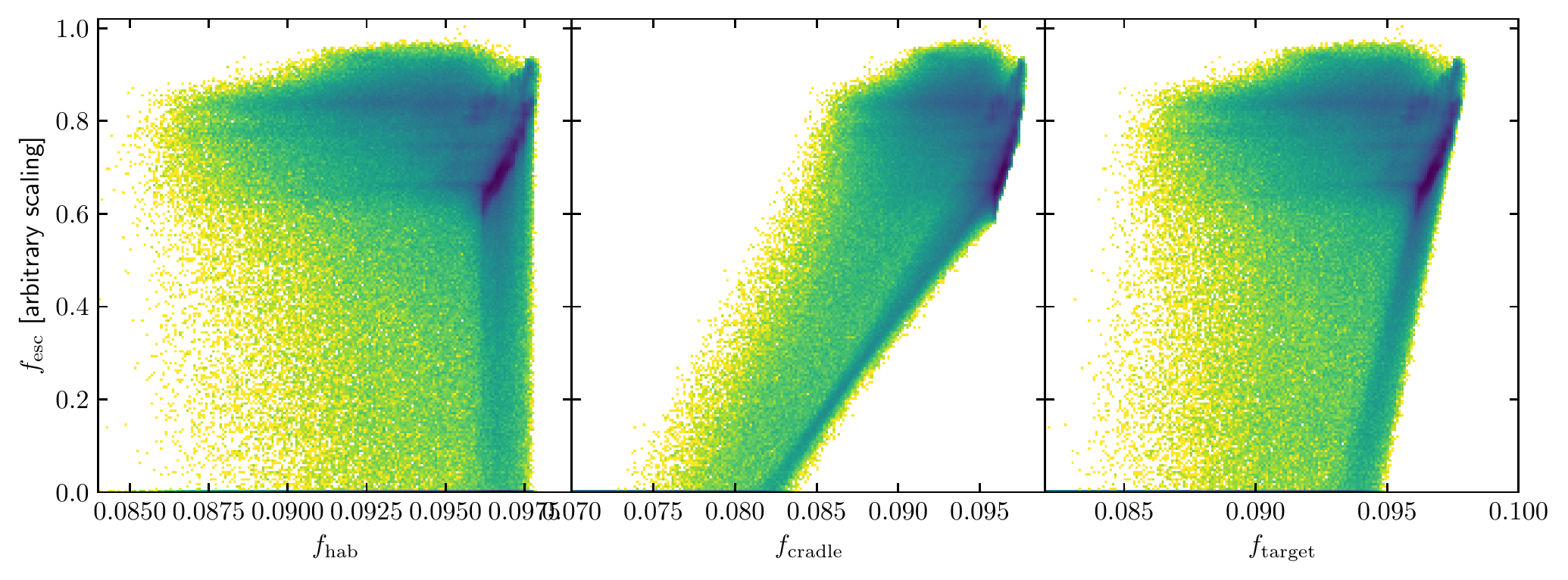}
\caption{The escape weight ($f_{\rm esc}$) as a function of three habitability levels ($f_{\rm hab}$, $f_{\rm cradle}$, and $f_{\rm target}$) in {\tt g15784} at $z = 0$, drawn in a logarithmic scale.}
\label{fig:fhab_fesc}
\end{figure*}

In the case of natural panspermia, we assume that life first develops on the surface of a planet within its star HZ, and that spores are then ejected from it by high velocity impacts. In general, few fragments reach planetary escape velocity and none stellar escape velocity. For interstellar panspermia to happen, therefore, the fragments must be accelerated by unspecified processes, which we assume only depend on the host mass star through its escape velocity within the HZ. Fig.~\ref{fig:ejecta1} shows the distribution of fragment velocities after impact and the (arbitrarily-scaled) stellar escape fraction. Fig.~\ref{fig:ejecta2} shows the evolution of the escape fraction $w_{\rm esc}$~over time, assuming that the frequency of impacts follows the one derived from Lunar cratering \citep{Neukum1994}:
\begin{equation}\label{eq:nimpact}
N_{\rm impact}(t)\propto \exp \left(6.93\,\frac{t_{\rm E}-t}{\rm Gyr} \right)\, ,
\end{equation}
where $t_{\rm E}$ the current age of the Earth. Finally, Fig.~\ref{fig:fesc_radial} shows the distribution of escape fractions $f_{\rm esc}$~computed for star particles, as a function of their galactocentric distance, and Fig.~\ref{fig:fhab_fesc} their correlation with habitability.
There exists a strong correlation $f_{\rm esc}$ and habitability within the {\tt DiskHalo} region, whereas star particles in the {\tt CentralDisk} tend to have lower $w_{\rm esc}$ due to their higher metallicities.
Interestingly, rather sharp linear lower boundaries exist in both $f_{\rm cradle}$-$f_{\rm esc}$ and $f_{\rm target}$-$f_{\rm esc}$ parameter spaces, where most of $f_{\rm esc}$ goes to zero at $f_{\rm cradle} < 0.082$ and $f_{\rm target} < 0.095$ (Fig.~\ref{fig:fhab_fesc}).
Finally, the {\tt Spheroids} show a radial distribution of $f_{\rm esc}$ similar to that of the main galaxy, though at smaller scale. 

\section{Unnormalized panspermia probabilities}\label{appendix:fhat}

\begin{figure}[h]
\centering
\plotone{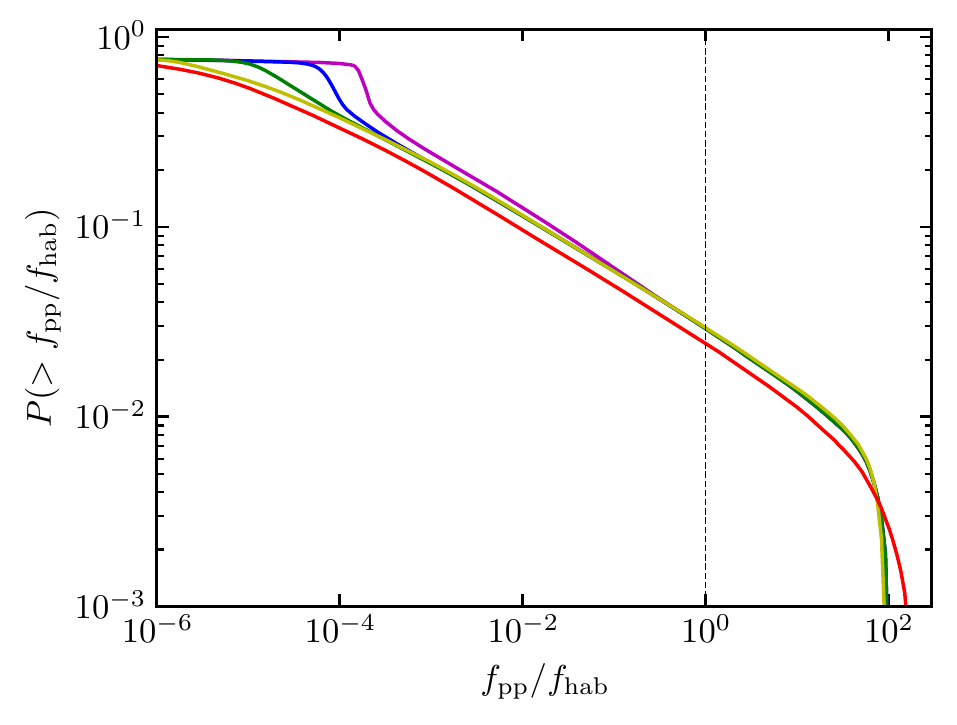}\\
\plotone{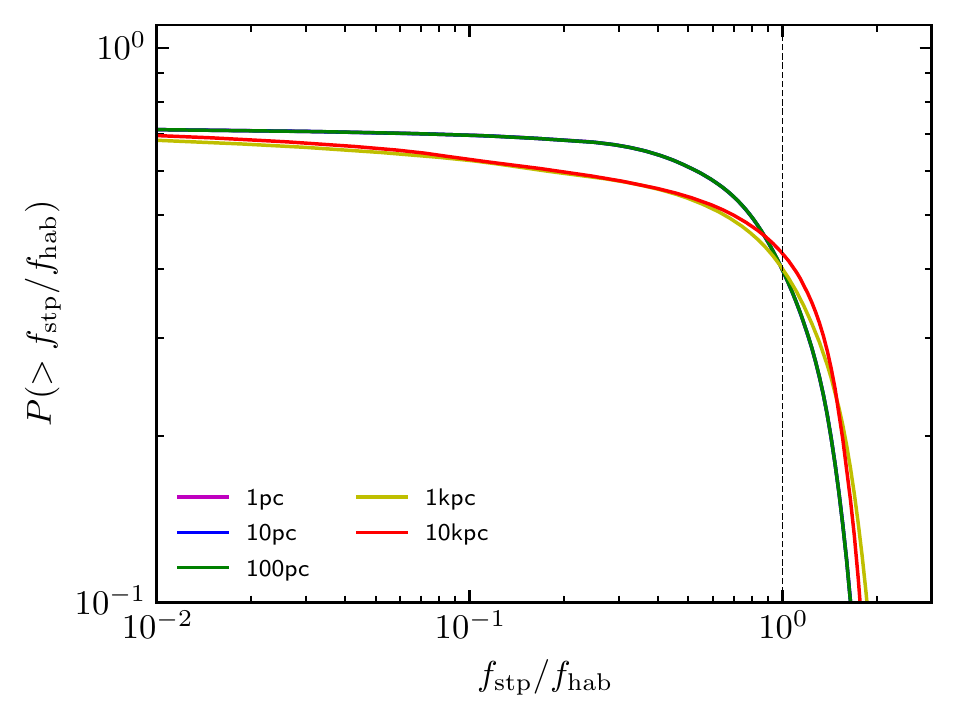}
\caption{Cumulative probabilities of the ratios between panspermia probabilities ($f_{\rm pp/stp}$) and natural habitability ($f_{\rm hab}$) in {\tt g15784} at $z = 0$, assuming the {\tt exp} damage model with varying $\ell_{\rm surv}$. The dashed vertical line marks $f_{\rm pp/stp} = f_{\rm hab}$. Both $f_{\rm pp}$ and $f_{\rm stp}$ are normalized so that their sum over {\tt g15784} is same as the sum of $f_{\rm hab}$.}
\label{fig:ratio_prob}
\end{figure}

Fig.~\ref{fig:ratio_prob} shows the cumulative probabilities of $f_{\rm pp/stp} / f_{\rm hab}$ of {\tt g15784} at $z = 0$. In both cases, the overall shape of the probability distribution is similar to those of $f_{\rm pp/stp}$, mainly due to the narrow range of $f_{\rm hab}$. Specifically, $P( > f_{\rm pp}/f_{\rm hab})$ follows a power-law relation of $P \propto (f_{\rm pp}/f_{\rm hab})^{-1/4}$ for $10^{-3} \lesssim f_{\rm pp} / f_{\rm hab} \lesssim 10^2$, falling to (nearly) zero for $f_{\rm pp} / f_{\rm hab} \gtrsim 10^2$. On the other hand, $P( > f_{\rm stp} / f_{\rm hab})$ is nearly zero for $f_{\rm stp} / f_{\rm hab} \gtrsim 2$ and plateaus at $P \simeq 0.7$ around $f_{\rm stp} / f_{\rm hab} \simeq 0.3$ and $0.01$ for $\ell_{\rm surv} \lesssim 100 \pc$ and $\gtrsim 100\pc$, respectively.
We note that the cumulative probability does not exceed 70\% in {\tt g15784} because the remaining 30\% of star particles have zero habitabilites.

Table~\ref{tab:ratio_prob} shows probabilities of $\hat{f}_{\rm pp/stp} > f_{\rm hab}$ for various values $\mathcal{R}$. The probability of star particles with $\hat{f}_{\rm pp} > f_{\rm hab}$ is much lower than the na\"ive expectation that $P(\mathcal{R}) \simeq \mathcal{R} / (\mathcal{R} + 1)$. For example, when $\mathcal{R} = 1$, i.e., the total panspermia probability and total habitability is the same, only 3\% of star particles have $\hat{f}_{\rm pp} > f_{\rm hab}$, rather than half as would be expected. In fact, $P(\hat{f}_{\rm pp} > f_{\rm hab}) > 1/2$ occurs only at very high values of $\mathcal{R} \gtrsim 10^{3.5}$. This ensures that only a tiny fraction of star particles dominate the panspermia process of the entire galaxy. On the other hand, $P(\hat{f}_{\rm stp} > f_{\rm hab})$ matches better the na\"ive expectation for values of $\mathcal{R} \gtrsim 1$.

\begin{deluxetable}{CCC}[bt]
\tablecaption{Probability of star particles with \emph{unnormalized} panspermia probabilities greater than natural habitability in {\tt g15784} at $z = 0$, for various values of $\mathcal{R}$).}\label{tab:ratio_prob}
\tablewidth{0.4\textwidth}
\tablehead {
\colhead{\hspace*{15pt}$\mathcal{R}$\hspace*{15pt}} &
\colhead{\hspace*{15pt}$P(\hat{f}_{\rm pp} > f_{\rm hab})$\hspace*{15pt}} &
\colhead{\hspace*{15pt}$P(\hat{f}_{\rm stp} > f_{\rm hab})$\hspace*{15pt}}
}
\startdata
$10^{-2}$ & $\lesssim 0.1\%$ & 0 \\
$10^{-1}$ & 1\% &  0 \\
$10^0$ & 3\% & 40\% \\
$10^1$ & 6\% & 70\% \\
$10^2$ & 10\% & 70\% \\
$10^3$ & 20\% & 70\% \\
$10^4$ & 70\% & 70\% \\
\enddata
\end{deluxetable}

\end{document}